\documentclass[
	amsmath,
	a4paper,
	pre,		
	reprint,	
	fleqn,
	showpacs,
	floatfix,
	superscriptaddress
]{revtex4-1}

\usepackage[bitstream-charter,greekuppercase=italicized]{mathdesign}
\usepackage[T1]{fontenc}

\usepackage[utf8]{inputenc}
\usepackage{microtype}

\usepackage{amssymb}

\DeclareSymbolFont{cmcal}{OMS}{cmsy}{m}{n}
\SetSymbolFont{cmcal}{bold}{OMS}{cmsy}{b}{n}
\DeclareSymbolFontAlphabet{\mathcal}{cmcal}
\DeclareMathAlphabet{\mathsf}{OT1}{iwona}{m}{n} 

\usepackage{tabularx}
\usepackage{booktabs}
\usepackage{multirow}

\usepackage{bm}
\usepackage{amsthm}
\usepackage{braket}
\usepackage{kbordermatrix}
	\renewcommand{\kbldelim}{(}
	\renewcommand{\kbrdelim}{)}

\usepackage{graphicx}
\usepackage{dcolumn}
\usepackage[caption=false]{subfig}
\usepackage{color}

\usepackage{verse}

\usepackage{hyperref}

\newcommand{\relation}{theorem}

\newcommand{\relations}{theorems}
\newcommand{\Relations}{Theorems}

\newcommand{\myTitle}{Detailed Fluctuation \Relations: A Unifying Perspective}

\newcommand{\myName}{Riccardo Rao}

\newcommand{\myAffiliation}{Complex Systems and Statistical Mechanics, Physics and Materials Science Research Unit, University of Luxembourg, L-1511 Luxembourg, G.D.~Luxembourg}

\newcommand{\myAdvisor}{Massimiliano Esposito}

\newcommand{\kavliInst}{Kavli Institute for Theoretical Physics, University of California, Santa Barbara, CA 93106 Santa Barbara,  U.S.A.}

\newcommand{\myFunding}
{
	This work was funded by the \emph{Luxembourg National Research Fund} (AFR PhD Grant 2014-2, No.~9114110), the \emph{European Research Council} (project NanoThermo, ERC-2015-CoG Agreement No.~681456), and the \emph{National Science Foundation} (NSF Grant No.~PHY-1748958).
}

\newcommand{\de}{\mathrm{d}}
\newcommand{\dt}{\mathrm{d}_{t}}

\newcommand{\at}[2]{\left.{#1}\right|_{#2}}

\newcommand{\ave}[1]{\left\langle {#1} \right\rangle}
\newcommand{\avef}[1]{\langle {#1} \rangle}

\newcommand{\transpose}{^{\mathsf{T}}}

\newcommand{\kb}{k_\mathrm{B}}

\newcommand{\ep}{\Sigma}
\newcommand{\epr}{\avef{\dot{\ep}}}
\newcommand{\vep}[1]{\ep_{\mathrm{#1}}}
\newcommand{\vepr}[1]{\avef{\dot{\ep}_{\mathrm{#1}}}}

\DeclareMathOperator{\diag}{diag}

\newcommand{\nof}[1]{\mathsf{N}_{#1}}

\definecolor{butter1}{rgb}{0.98,0.91,0.31}
\definecolor{butter2}{rgb}{0.93,0.83,0}
\definecolor{butter3}{rgb}{0.77,0.63,0}
\definecolor{skyblue1}{rgb}{0.45,0.62,0.81}
\definecolor{skyblue2}{rgb}{0.2,0.39,0.64}
\definecolor{skyblue3}{rgb}{0.13,0.29,0.53}
\definecolor{scarlet1}{rgb}{0.93,0.16,0.16}
\definecolor{scarlet2}{rgb}{0.8,0,0}
\definecolor{scarlet3}{rgb}{0.64,0,0}
\definecolor{chameleon1}{rgb}{0.54,0.88,0.2}
\definecolor{chameleon2}{rgb}{0.45,0.82,0.09}
\definecolor{chameleon3}{rgb}{0.3,0.6,0.02}
\definecolor{orange1}{rgb}{0.98,0.68,0.24}
\definecolor{orange2}{rgb}{0.96,0.47,0}
\definecolor{orange3}{rgb}{0.8,0.36,0}
\definecolor{plum1}{rgb}{0.68,0.5,0.66}
\definecolor{plum2}{rgb}{0.46,0.31,0.48}
\definecolor{plum3}{rgb}{0.36,0.21,0.4}
\definecolor{chocolate1}{rgb}{0.91,0.72,0.43}
\definecolor{chocolate2}{rgb}{0.75,0.49,0.07}
\definecolor{chocolate3}{rgb}{0.56,0.35,0.01}
\definecolor{aluminium1}{rgb}{0.93,0.93,0.92}
\definecolor{aluminium2}{rgb}{0.82,0.84,0.81}
\definecolor{aluminium3}{rgb}{0.73,0.74,0.71}
\definecolor{aluminium4}{rgb}{0.53,0.54,0.52}
\definecolor{aluminium5}{rgb}{0.33,0.34,0.32}
\definecolor{aluminium6}{rgb}{0.18,0.2,0.21}  

\newcommand{\greyt}[1]{\textcolor{aluminium5}{#1}}

\definecolor{webgreen}{rgb}{0,.5,0}
\definecolor{webbrown}{rgb}{.6,0,0}
\definecolor{grigio}{rgb}{.85,.85,.85} 
\definecolor{RoyalBlue}{rgb}{0.0, 0.14, 0.4}

\hypersetup{%
    colorlinks=true, linktocpage=true, pdfstartpage=1, pdfstartview=FitV,%
    breaklinks=true, pdfpagemode=UseNone, pageanchor=true, pdfpagemode=UseOutlines,%
    plainpages=false, bookmarksnumbered, bookmarksopen=true, bookmarksopenlevel=1,%
    hypertexnames=true, pdfhighlight=/O,
    urlcolor=webbrown, linkcolor=RoyalBlue, citecolor=webgreen, 
    pdftitle={\myTitle},%
    pdfauthor={\textcopyright\ \myName},%
    pdfsubject={},%
    pdfkeywords={},%
    pdfcreator={pdfLaTeX},%
}

\newcommand{\yp}{y_{\mathrm{p}}}
\newcommand{\yf}{y_{\mathrm{f}}}

\newcommand{\lX}{\kappa}
\newcommand{\X}{X^{\lX}}
\newcommand{\deltaX}{\delta X}

\newcommand{\trj}{\bm n_{t}}
\newcommand{\prt}{\pi_{t}}

\renewcommand{\nof}[1]{\mathsf{N}_{#1}}
\renewcommand{\forall}{\text{for all }}

\newcommand{\pref}[1]{p^{\mathrm{ref}}_{#1}}
\newcommand{\Aref}[1]{A^{\mathrm{ref}}_{#1e}}
\newcommand{\Pref}[1]{\psi^{\mathrm{ref}}_{#1}}

\newcommand{\p}[2]{p^{\mathrm{#1}}_{#2}}
\newcommand{\A}[2]{A^{\mathrm{#1}}_{#2e}}
\renewcommand{\P}[2]{\psi^{\mathrm{#1}}_{#2}}
\newcommand{\pss}[1]{p^{\mathrm{ss}}_{#1}}
\newcommand{\gauge}[1]{\P{}{#1}}

\newcommand{\ST}{\mathcal{T}}
\newcommand{\chords}{\mathcal{T}^{\ast}}

\newcommand{\cycle}{\mathcal{C}}

\newcommand{\edge}{\mathcal{E}}

\renewcommand{\o}[1]{\mathfrak{o}(#1)}
\renewcommand{\t}[1]{\mathfrak{t}(#1)}

\newcommand{\obs}{O}
\newcommand{\deltaO}{\deltaup O}
\newcommand{\cf}{q}

\begin{document}

\title{\myTitle}

\author{\myName}
\affiliation{\myAffiliation}
\author{\myAdvisor}
\affiliation{\myAffiliation}
\affiliation{\kavliInst}

\date{\today. Published in \emph{Entropy}, DOI:~\href{https://doi.org/10.3390/e20090635}{10.3390/e20090635}}

\begin{abstract}
	We present a general method to identify an arbitrary number of fluctuating quantities which satisfy a detailed fluctuation theorem for all times within the framework of time-inhomogeneous Markovian jump processes.
	In doing so we provide a unified perspective on many fluctuation theorems derived in the literature.
	By complementing the stochastic dynamics with a thermodynamic structure (\emph{i.e.} using stochastic thermodynamics), we also express these fluctuating quantities in terms of physical observables.
\end{abstract}


\maketitle

\section{Introduction}

The discovery of different fluctuation \relations{} (FTs) over the last two decades constitutes a major progress in nonequilibrium physics \citep{harris07,esposito09,jarzynski11,campisi11,seifert12,vandenbroeck15}.
These relations are exact constraints that some fluctuating quantities satisfy arbitrarily far from equilibrium. 
They have been verified experimentally in many different contexts, ranging from biophysics to electronic circuits \cite{ciliberto17}.
However, they come in different forms: detailed fluctuation \relations{} (DFTs) or integral fluctuation \relations{} (IFTs), and concern various types of quantities.
Understanding how they are related and to what extent they involve mathematical quantities or interesting physical observables can be challenging.
 
The aim of this paper is to provide a simple yet elegant method to identify a class of finite-time DFTs for time-inhomogeneous Markovian jump processes.
The method is based on splitting the entropy production (EP) in three contributions by introducing a reference probability mass function (PMF).
The latter is parametrized by the time-dependent driving protocol, which renders the dynamics time-inhomogeneous.
The first contribution quantifies the EP as if the system were in the reference PMF, the second the extent to which the reference PMF changes with the driving protocol, and the last the mismatch between the actual and the reference PMF.
We show that when the system is initially prepared in the reference PMF, the joint probability distribution for the first two terms always satisfies a DFT.
We then show that various known DFTs can be immediately recovered as special cases.
We emphasize at which level our results make contact with physics and also clarify the nontrivial connection between DFTs and EP fluctuations.
Our EP splitting is also shown to be connected to information theory.
Indeed, it can be used to derive a generalized Landauer principle identifying the minimal cost needed to move the actual PMF away from the reference PMF.
While unifying, we emphasize that our approach by no means encompasses all previously derived FTs and that other FT generalizations have been made (\emph{e.g.},~\cite{chetrite11,seifert12,perez-espigares12,verley12,baiesi15}).

The plan of this paper is as follows.
Time-inhomogeneous Markov jump processes are introduced in Sec.~\ref{sec:jumpProc}.
Our main results are presented in Sec.~\ref{sec:GenDec}:
We first introduce the EP as a quantifier of detailed balance breaking, and we then show that by choosing a reference PMF, a splitting of the EP ensues.
This enables us to identify the fluctuating quantities satisfying a DFT and an IFT when the system is initially prepared in the reference PMF.
While IFTs hold for arbitrary reference PMFs, DFTs require reference PMFs to be solely determined by the driving protocol encoding the time dependence of the rates.
The EP decomposition is also shown to lead to a generalized Landauer principle.
The remaining sections are devoted to selecting specific reference PMFs and showing that they give rise to interesting mathematics or physics:
In Sec.~\ref{sec:sse} the steady-state PMF of the Markov jump process is chosen, giving rise to the adiabatic--nonadiabatic split of the EP \cite{esposito07}.
In Sec.~\ref{sec:ste} the equilibrium PMF of a spanning tree of the graph defined by the Markov jump process is chosen, and gives rise to a cycle--cocycle decomposition of the EP \cite{polettini14:cocycle}.
Physics is introduced in Sec.~\ref{stochTherm}, and the properties that the Markov jump process must satisfy to describe the thermodynamics of an open system are described.
In Sec.~\ref{sec:mce} the microcanonical distribution is chosen as the reference PMF, leading to the splitting of the EP into system and reservoir entropy change.
Finally, in Sec.~\ref{sec:gge}, the generalized Gibbs equilibrium PMF is chosen as a reference and leads to a {conservative--nonconservative} splitting of the EP \cite{rao18:shape}.
Conclusions are finally drawn, and some technical proofs are discussed in the appendices.

\section{Markov Jump Process}
\label{sec:jumpProc}

We introduce time-inhomogeneous Markovian jump processes and set the notation.

We consider an externally driven open system described by a finite number of states, which we label by $n$.
Allowed transitions between pairs of states are identified by directed edges,
\begin{equation}
	e \equiv (nm, \nu) \, , \quad \text{for } n \overset{\nu}{\longleftarrow } m \, ,
	\label{}
\end{equation}
where the label $\nu$ indexes different transitions between the same pair of states (\emph{e.g.}, transitions due to different reservoirs).
The evolution in time of the probability of finding the system in the state $n$, $p_{n} \equiv p_{n}(t)$, is ruled by the \emph{master equation} (ME):
\begin{equation}
	\dt p_{n} = {\textstyle\sum_{m}} W_{nm} p_{m} \, ,
	\label{eq:MEgenerator}
\end{equation}
where the elements of the \emph{rate matrix} are represented as
\begin{equation}
	W_{nm} = {\textstyle\sum_{e}} w_{e} \left\{ \delta_{n,\t{e}} \delta_{m,\o{e}} - \delta_{n,m} \delta_{m,\o{e}} \right\} \, .
	\label{eq:generator}
\end{equation}
The latter is written in terms of stochastic transition rates, $\set{w_{e}}$, and the functions
\begin{equation}
	\o{e} := m \, , \; \text{and } \quad \t{e}:= n \, , \quad \text{for } e = (nm, \nu) \, ,
	\label{eq:origin}
\end{equation}
which map each transition to the state from which it originates (origin) and to which it leads (target), respectively.
The off-diagonal entries of the rate matrix (the first term in brackets) give the probability per unit time to transition from $m$ to $n$.
The diagonal ones (second term in brackets) are the escape rates denoting the probability per unit time of leaving the state $m$.
For thermodynamic consistency, we assume that each transition $e \equiv (nm, \nu)$ is reversible, namely if $w_{e}$ is finite, the corresponding backward transition $-e \equiv (mn, \nu)$ is allowed and additionally  has a finite rate $w_{-e}$.
For simplicity, we also assume that the rate matrix is irreducible at all times, so that the stochastic dynamics is ensured to be ergodic.
The Markov jump process is said to be \emph{time-inhomogeneous} when the transition rates depend on time.
The driving \emph{protocol value} $\pi_{t}$ determines the values of all rates at time $t$, $\set{w_{e} \equiv w_{e}(\pi_{t})}$.

The ME~\eqref{eq:MEgenerator} can be rewritten as a continuity equation: 
\begin{equation}
	\dt p_{n} = {\textstyle\sum_{e}} D^{n}_{e} \, \avef{j^{e}} \, ,
	\label{eq:ME}
\end{equation}
where we introduced the averaged transition \emph{probability fluxes},
\begin{equation}
	\avef{j^{e}} = w_{e} p_{\o{e}} \, ,
	\label{eq:aveCurrents}
\end{equation}
and the \emph{incidence matrix} $D$,
\begin{equation}
	D^{n}_{e} :=
	\delta_{n,\t{e}} - \delta_{n,\o{e}}
	=
	\begin{cases}
		+ 1 & \text{if } \overset{e}{\longrightarrow} n \, , \\
		- 1 & \text{if } \overset{e}{\longleftarrow} n \, , \\
		  0 & \text{otherwise} \, ,
	\end{cases}
	\label{eq:incidence}
\end{equation}
which couples each transition to the pair of states that it connects, and hence encodes the \emph{network topology}.
On the graph identified by the vertices $\set{n}$ and the edges $\set{e}$, it can be viewed as a (negative) divergence operator when acting on edge-space vectors---as in the ME \eqref{eq:ME}---or as a gradient operator when acting on vertex-space vectors.
It satisfies the symmetry $D^{n}_{-e} = - D^{n}_{e}$.

\begin{figure}[t]
	\centering
	\includegraphics[width=.25\textwidth]{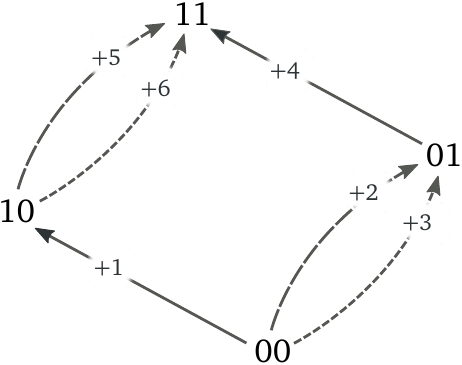}
	\caption{Illustration of a network of transitions.}
	\label{fig:network}
\end{figure}

\paragraph*{Example}
Let us consider the Markov jump process on the network in Fig.~\ref{fig:network}, in which only the six forward transitions are depicted.
It is characterized by four states, $\set{00,01,10,11}$, connected by transitions as described by the incidence matrix:
\begin{equation}
	D =
	\kbordermatrix{
		& \greyt{+1} & \greyt{+2} & \greyt{+3} & \greyt{+4} & \greyt{+5} & \greyt{+6} \\
		\greyt{00} & -1 & -1 & -1 & 0 & 0 & 0 \\
		\greyt{10} & 1 & 0 & 0 & 0 & -1 & -1 \\
		\greyt{01} & 0 & 1 & 1 & -1 & 0 & 0 \\
		\greyt{11} & 0 & 0 & 0 & 1 & 1 & 1 
	} \, .
	\label{}
\end{equation}
Backward transitions are obtained from $D^{n}_{-e} = - D^{n}_{e}$.

\paragraph*{Notation}
From now on, upper--lower indices and Einstein summation notation will be used:
repeated upper--lower indices implies the summation over all the allowed values for those indices.
Time derivatives are denoted by ``$\dt$'' or ``$\partial_{t}$'', whereas the overdot ``$\; \dot{} \;$'' is reserved for rates of change of quantities that are not exact time derivatives of state functions.
We also take the Boltzmann constant $\kb$ equal to $1$.

\section{General Results}
\label{sec:GenDec}

This section constitutes the core of the paper.
The main results are presented in their most general form. 

\subsection{EP Decomposition at the Ensemble Average Level}

After defining the ensemble-averaged EP, we will show how to generically decompose it in terms of a reference PMF.

A PMF $p_{n}$ satisfies the \emph{detailed-balance} property if and only if
\begin{equation}
	w_{e} p_{\o{e}} = w_{-e} p_{\o{-e}} \, , \quad \text{for all transitions } e \, .
	\label{eq:db}
\end{equation}
This implies that all net transition probability currents vanish: $\avef{j^{e}} - \avef{j^{-e}} = 0$.
The central quantity that we will consider is the \emph{EP rate}:
\begin{equation}
	\epr = \tfrac{1}{2} \, A_{e} \left[ \avef{j^{e}} - \avef{j^{-e}} \right] = A_{e} \avef{j^{e}} \ge 0 \, ,
	\label{eq:epr}
\end{equation}
where the \emph{affinities} are given by
\begin{equation}
	A_{e}
	= \ln \frac{w_{e} p_{\o{e}}}{w_{-e} p_{\o{-e}}} \, .
	\label{eq:affinity}
\end{equation}
It is a measure of the amount by which the system breaks detailed balance or, equivalently, time-reversal symmetry.
Indeed, its form ensures that it is always non-negative and vanishes if and only if Eq.~\eqref{eq:db} holds.
Notice that $A_{-e} = - A_{e}$.
As we will see in Sec.~\ref{sec:mce}, in physical systems the EP quantifies the total entropy change in the system plus environment \cite{schnakenberg76}.

We now decompose the EP rate into two contributions using a generic PMF $\pref{n} \equiv \pref{n}(t)$ as a \emph{reference}.
We make no assumption about the properties of $\pref{n}$ at this stage, and define the reference potential and the reference affinities as
\begin{equation}
	\Pref{n} := - \ln \pref{n}
	\label{eq:Pref}
\end{equation}
and
\begin{equation}
	\Aref{} := \ln \frac{w_{e} \pref{\o{e}}}{w_{-e} \pref{\o{-e}}} = \ln \frac{w_{e}}{w_{-e}} + \Pref{n} D^{n}_{e} \, ,
	\label{eq:Aref}
\end{equation}
respectively.
The former can be thought of as the entropy associated to $\pref{n}$---\emph{i.e.}, its \emph{self-information}---, whereas the latter measures the extent by which $\pref{n}$ breaks detailed balance.
By merely adding and subtracting $\Pref{n} D^{n}_{e}$ from the EP rate, the latter can be formally decomposed as
\begin{equation}
	\epr = \vepr{nc} + \vepr{c} \geq 0 \, ,
	\label{eq:eprMaster}
\end{equation}
where the \emph{reference nonconservative contribution} is an EP with affinities replaced by reference affinities: 
\begin{equation}
	\vepr{nc} := \Aref{} \avef{j^{e}} \, ,
	\label{eq:eprMasterNC}
\end{equation}
and the \emph{reference conservative contribution} is
\begin{equation}
	\vepr{c} := - {\textstyle\sum_n} \dt p_{n} \ln \left\{ p_{n} / \pref{n} \right\} \, .
	\label{eq:eprMasterC}
\end{equation}
Using the ME \eqref{eq:ME}, it can be further decomposed as
\begin{equation}
	\vepr{c} = - \dt \mathcal{D}(p\|\pref{}) + \vepr{d} \, ,
	\label{eq:eprMasterCD}
\end{equation}
where the first term quantifies the change in time of the \emph{dissimilarity} between $p_{n}$ and $\pref{n}$, since
\begin{equation}
	\mathcal{D}(p\|\pref{}) := {\textstyle\sum_{n}} p_{n} \ln \left\{ p_{n} / \pref{n} \right\}
	\label{eq:relativeEntropy}
\end{equation}
is a \emph{relative entropy}, whereas the second term,
\begin{equation}
	\vepr{d} := - {\textstyle\sum_{n}} p_{n} \dt \ln \pref{n} = {\textstyle\sum_{n}} p_{n} \dt \Pref{n} \, ,
	\label{eq:eprMasterD}
\end{equation}
accounts for possible time-dependent changes of the reference state, and we name it the \emph{driving contribution}.
The reason for this name will become clear later, as we will request $\pref{n}$ to depend parametrically on time only via the driving protocol (\emph{i.e.}, $\pref{n}(t) = \pref{n}(\pi_{t})$).

Using these equations, one can easily rearrange Eq.~\eqref{eq:eprMaster} into 
\begin{equation}
	\vepr{d} + \vepr{nc} \geq \dt \mathcal{D}(p\|\pref{}) \, .
	\label{eq:eprLandauer}
\end{equation}
When $\pref{n}(t) = \pref{n}(\pi_{t})$, one can interpret this equation as follows.
The lhs describes the EP contribution due to the time-dependent protocol, $\vepr{d}$, and to the break of detailed balance required to sustain the reference PMF, $\vepr{nc}$.
When positive, the rhs thus represents the minimal cost (ideally achieved at vanishing EP) to move the PMF further away from the reference PMF.
When negative, its absolute value becomes the maximal amount by which the two EP contributions can decrease, as the PMF approaches the reference PMF.
This result can be seen as a \emph{mathematical generalization of the Landauer principle}, as it provides a connection between an information-theoretical measure of the dissimilarity between two PMFs and the driving and break of detailed balance needed to achieve it. 
Its precise physical formulation, discussed in detail in \cite{rao18:shape}, is obtained when expressing Eq.~\eqref{eq:eprLandauer} in terms of the reference PMF used in Sec.~\ref{sec:gge}.

\subsection{EP Decomposition at the Trajectory Level}

We now perform the analogue of the EP decomposition~\eqref{eq:eprMaster} at the level of single stochastic trajectories.

A stochastic \emph{trajectory} of duration $t$, $\trj$, is defined as a set of transitions $\{e_{i}\}$ sequentially occurring at times $\{t_{i}\}$ starting from $n_{0}$ at time $0$.
If not stated otherwise, the transitions index $i$ runs from $i=1$ to the last transition prior to time $t$, $\nof{t}$, whereas the state at time $\tau \in [0,t]$ is denoted by $n_{\tau}$.
The whole trajectory is encoded in the \emph{instantaneous fluxes},
\begin{equation}
	j^{e}(\tau) := {\textstyle\sum_{i}} \delta_{e,e_{i}} \delta(\tau - t_{i}) \, ,
	\label{eq:instantaneousCurrent}
\end{equation}
as they encode the transitions that occur and their timing.
Its corresponding trajectory probability measure is given by
\begin{multline}
	\mathfrak{P}[\trj;\prt] = \prod_{i=1}^{\nof{t}} w_{e_{i}}(\pi_{t_{i}}) \\
	\prod_{i=0}^{\nof{t}} \exp\left\{ - {\textstyle\int_{t_{i}}^{t_{i+1}}} \de \tau \, {\textstyle\sum_{e}} w_{e}(\pi_{\tau}) \delta_{n_{\tau},\o{e}} \right\} \, ,
	\label{eq:probTrajectory}
\end{multline}
where the first term accounts for the probability of transitioning along the edges, while the second accounts for the probability that the system spends $\set{t_{i+1} - t_{i}}$ time in the state $\set{n_{t_{i}}}$.
When averaging Eq.~\eqref{eq:instantaneousCurrent} over all stochastic trajectories, we obtain the averaged fluxes, Eq.~\eqref{eq:aveCurrents},
\begin{equation}
	\avef{j^{e}(\tau)}
	= {\textstyle\int}\mathfrak{D}\trj \, \mathfrak{P}[\trj;\prt] \, p_{n_{0}}(0) \, j^{e}(\tau) \, ,
	\label{eq:aveCurrentGivenState}
\end{equation}
where ${\textstyle\int}\mathfrak{D}\trj$ denotes the integration over all stochastic trajectories.

The change along $\trj$ of a state function like $\Pref{n}$ can be expressed as
\begin{equation}
	\hspace{-1.6em}
	\begin{split}
		\Delta & \Pref{}[\trj] = \Pref{n_{t}}(t) - \Pref{n_{0}}(0) \\
		& = {\int_{0}^{t}} \de \tau \left\{ \at{\left[ \de_{\tau} \Pref{n}(\tau) \right]}{n=n_{\tau}} + \Pref{n}(\tau) \, D^{n}_{e} \, j^{e}(\tau) \right\} \, .
	\end{split}
	\label{eq:PrefTrj}
\end{equation}
The first term on the rhs accounts for the instantaneous changes of $\pref{n}$, while the second accounts for its finite changes due to stochastic transitions.
Analogously, the trajectory EP---which is not a state function---can be written as
\begin{equation}
	\Sigma[\trj;\prt] = {\int_{0}^{t}} \de \tau \, j^{e}(\tau) \ln \frac{w_{e}(\pi_{\tau})}{w_{-e}(\pi_{\tau})} - \ln \frac{p_{n_{t}}(t)}{p_{n_{0}}(0)} \, .
	\label{eq:ep}
\end{equation}
Adding and subtracting the terms of Eq.~\eqref{eq:PrefTrj} from the EP, we readily obtain the fluctuating expressions of the nonconservative and conservative contributions of the EP,
\begin{equation}
	\ep[\trj;\prt] = \vep{nc}[\trj;\prt] + \vep{c}[\trj] \, .
	\label{eq:epMaster}
\end{equation}
The former reads
\begin{equation}
	\vep{nc}[\trj;\prt]
	= {\textstyle\int_{0}^{t}} \de \tau \, \Aref{}(\tau) \, j^{e}(\tau) \, ,
	\label{eq:epMasterNC}
\end{equation}
while for the latter
\begin{equation}
	\vep{c}[\trj] = - \Delta \mathcal{D}[\trj] + \vep{d}[\trj] \, ,
	\label{eq:epMasterC}
\end{equation}
where
\begin{equation}
	\Delta \mathcal{D}[\trj] := \ln \frac{p_{n_{t}}(t)}{\pref{n_{t}}(t)} - \ln \frac{p_{n_{0}}(0)}{\pref{n_{0}}(0)}
	\label{eq:epMasterB}
\end{equation}
and
\begin{equation}
	\vep{d}[\trj] := {\int_{0}^{t}} \de \tau \at{\left[ \de_{\tau} \Pref{n}(\tau) \right]}{n=n_{\tau}} \, .
	\label{eq:epMasterD}
\end{equation}

We emphasize that Eq.~\eqref{eq:epMaster} holds for any reference PMF $\pref{n}$ exactly as it was for its ensemble-averaged rate counterpart, Eq.~\eqref{eq:eprMaster}.

\subsection{Fluctuation \Relations}

\begin{figure}[t]
	\centering
	\includegraphics[width=.45\textwidth]{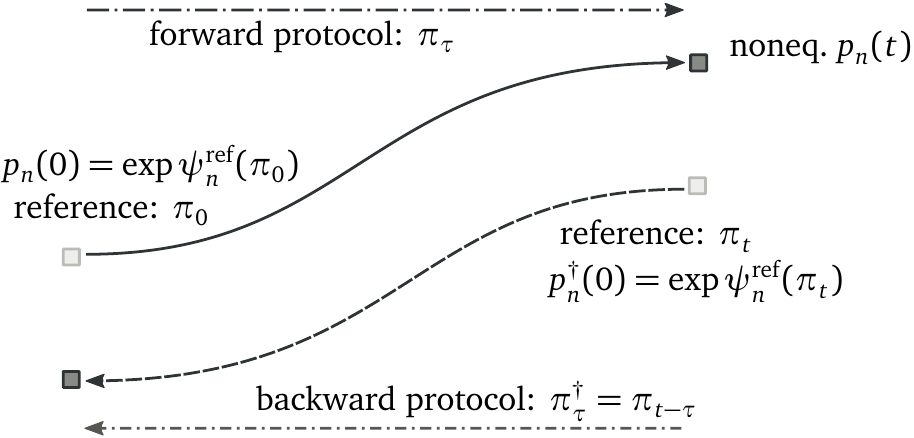}
	\caption{
		Schematic representation of the forward and backward processes related by our detailed fluctuation \relation{} (DFT).
	}
	\label{fig:fr}
\end{figure}

We proceed to show that a class of FTs ensue from the decomposition \eqref{eq:eprMaster}--\eqref{eq:epMaster}.
To do so, we now need to assume that the reference PMF depends instantaneously \emph{solely} on the protocol value $\pref{n}(\tau) = \pref{n}(\pi_{\tau})$.
In other words, $\pref{n}$ at time $\tau$ is completely determined by $\set{w_{e}(\pi_{\tau})}$.
This justifies {a posteriori} the name driving contribution for Eq.~\eqref{eq:eprMasterD}.
Various instances of such PMFs will be provided in the following sections.
We define a \emph{forward process} where the system is initially prepared in $p_{n}(0) = \pref{n}(\pi_{0})$ at a value of the protocol $\pi_{0}$ and then evolves under the Markov jump process driven by a protocol $\pi_{\tau}$, for $\tau\in[0,t]$.
The corresponding \emph{backward process}, denoted with ``$\, ^{\dagger} \,$'', is defined as follows:
the system is initially prepared in the reference PMF corresponding to the final value of the forward process, $p^{\dagger}_{n}(0) = \pref{n}(\pi_{t})$, and then evolves under the Markov jump process driven by the forward protocol reversed in time,
\begin{equation}
	\pi^{\dagger}_{\tau} := \pi_{t - \tau} \, , \quad \text{for } \tau\in[0,t] \, ,
	\label{eq:trProt}
\end{equation}
see Fig.~\ref{fig:fr}.

Our main result is that the forward and backward process are related by the following \emph{finite-time DFT}:
\begin{equation}
	\frac{P_{t}(\vep{d},\vep{nc})}{P_{t}^{\dagger}(-\vep{d},-\vep{nc})} = \exp\left\{ \vep{d} + \vep{nc} \right\} \, .
	\label{eq:dftMaster}
\end{equation}
Here $P_{t}(\vep{d},\vep{nc})$ is the probability of observing a driving contribution to the EP $\vep{d}$ and a nonconservative one $\vep{nc}$ along the forward process.
Instead, $P_{t}^{\dagger}(-\vep{d},-\vep{nc})$ is the probability of observing a driving contribution equal to $-\vep{d}$, and a nonconservative one $-\vep{nc}$ along the backward process.

We now mention two direct implications of our DFT.
First, by marginalizing the joint probability, one easily verifies that the sum of nonconservative and driving EP contributions also satisfies a DFT:
\begin{equation}
	\frac{P_{t}(\vep{d}+\vep{nc})}{P_{t}^{\dagger}(-\vep{d}-\vep{nc})} = \exp\left\{ \vep{d} + \vep{nc} \right\} \, .
	\label{eq:dftMasterSum}
\end{equation}
Second, when averaging Eq.~\eqref{eq:dftMaster} over all possible values of $\vep{d}$ and $\vep{nc}$, an IFT ensues:
\begin{equation}
	\ave{\exp \left\{ - \vep{d} - \vep{nc} \right\}} = 1 \, .
	\label{eq:iftMaster}
\end{equation}
The proofs of Eqs.~\eqref{eq:dftMaster}--\eqref{eq:iftMaster} are given in App.~\ref{sec:ftMasterProof}, and use the generating function techniques developed in Refs.~\cite{rao18:shape,esposito07}.

We note that the IFT holds for any reference PMF regardless of the requirement that $\pref{n}(\tau) = \pref{n}(\pi_{\tau})$ (see App.~\ref{sec:ftMasterProof}).
In contrast, this requirement must hold for the DFT, else the probability $P^{\dagger}_{t}(\vep{d},\vep{nc})$ would no longer describe a physical backward process in which solely the protocol function is time reversed.
Indeed, if one considers an arbitrary $\pref{n}$, the backward process corresponds to not only reversing the protocol, but also the stochastic dynamics itself (see Eq.~\eqref{eq:trME}).

Another noteworthy observation is that the fluctuating quantity $\vep{d} + \vep{nc}$ can be seen as the ratio between the probabilities to observe a trajectory $\trj$ along the forward process, Eq.~\eqref{eq:probTrajectory}, and the probability to observe the time-reversed trajectory along the backward process:
\begin{equation}
	\hspace{-1em}
	\vep{nc}[\trj;\prt] + \vep{d}[\trj;\prt] = \ln \frac{\mathfrak{P}[\trj;\prt] \, \pref{n_{0}}(\pi_{0})}{\mathfrak{P}[\trj^{\dagger};\prt^{\dagger}] \, \pref{n_{t}}(\pi_{t})} \, .
	\label{eq:vepFluctuating}
\end{equation}
The latter trajectory is denoted by $\trj^{\dagger}$.
It starts from $n_{t}$, and it is defined by:
\begin{equation}
	{j^{\dagger\, e}}(\tau) := {\textstyle\sum_{i}} \delta_{e,-e_{i}} \delta(t-\tau-t_{i}) \, .
	\label{eq:trCrr}
\end{equation}
This result follows using Eq.~\eqref{eq:probTrajectory} and the observation that the contribution due to the waiting times vanish in the ratio on the rhs.
It can also be used to prove the DFT in two alternative ways, the first inspired by Ref.~\cite{garcia-garcia10} and the second using trajectory probabilities (see App.~\ref{sec:ftMasterProof2}).
These proofs rely on the fact that both the driving and the nonconservative EP contributions satisfy the \emph{involution} property:
\begin{equation}
	\begin{aligned}
		\vep{nc}[\trj^{\dagger};\prt^{\dagger}] & = - \vep{nc}[\trj;\prt] \, , \; \text{and} \\
		\vep{d}[\trj^{\dagger};\prt^{\dagger}] & = - \vep{d}[\trj;\prt] \, ,
	\end{aligned}
	\label{eq:involution}
\end{equation}
\emph{viz.} the change of $\vep{d}$ and $\vep{nc}$ for the backward trajectory along the backward process is minus the change along the forward trajectory of the forward process.
This result follows from direct calculation on Eqs.~\eqref{eq:epMasterNC} and \eqref{eq:epMasterD} (see App.~\ref{sec:ftMasterProof2}).

Finally, let us get back to the generalized Landauer principle for systems initially prepared in the reference state, as we did in this subsection for the FTs to hold.
Using Eq.~\eqref{eq:eprLandauer}, we see that the arguments of the FTs \eqref{eq:dftMasterSum} and \eqref{eq:iftMaster} (\emph{i.e.}, the driving and the nonconservative contribution to the EP) can be interpreted, on average, as the cost to generate a dissimilarity (or a lag) between the actual and the reference PMF at the end of the forward protocol.
A special case of this result is discussed in Ref.~\cite{vaikuntanathan09}. 

\subsection{EP Fluctuations}

We now discuss the properties of the fluctuating EP and its relation to the previously derived FTs.

An IFT for the EP always holds 
\begin{equation}
	\ave{\exp \left\{ - \ep \right\}} = 1 \, ,
	\label{eq:masterIFTEP}
\end{equation}
regardless of the initial condition \cite{seifert05}.
In our framework, this can be seen as the result of choosing the actual $p_{n}(\tau)$ as the reference for the IFT \eqref{eq:iftMaster}.

In contrast, a general DFT for the EP does not hold.
This can be easily understood at the level of trajectory probabilities.
Indeed, the fluctuating EP can be written as the ratio of forward and backward probabilities as in \eqref{eq:vepFluctuating}, but the initial condition of the forward process is arbitrary, and that of the backward process is the final PMF of the forward process,
\begin{equation}
	\ep[\trj;\prt] = \ln \frac{\mathfrak{P}[\trj;\prt] \, p_{n_{0}}(0)}{\mathfrak{P}[\trj^{\dagger};\prt^{\dagger}] \, p_{n_{t}}(t)} \, .
	\label{eq:epTrj}
\end{equation}
As a result, the involution property is generally lost, $\ep[\trj^{\dagger};\prt^{\dagger}] \neq - \ep[\trj;\prt]$, since $p^{\dagger}_{n_{0}}(t) \neq p_{n_{0}}(0)$, and hence the DFT is also lost \cite{seifert05}.

However, in special cases, the fluctuating quantity $\vep{d} + \vep{nc}$ which satisfies a DFT can be interpreted as an EP.
This happens if at the end of the forward (respectively backward) process, the protocol stops changing in time in such a way that the system relaxes from $p_{n_{t}}$ to an equilibrium $\pref{n_{t}}$ (respectively from $p^{\dagger}_{n}(t)$ to an equilibrium $\pref{n}(\pi_{0})$) and thus without contributing  to either $\vep{d}$ or to $\vep{nc}$ (even at the trajectory level).
In such cases, $\vep{d}+\vep{nc}$ can be seen as the EP of the \emph{extended} process including the relaxation.
On average, it is greater or equal than the EP of the same process without the relaxation, since the non-negative EP during the relaxation is given by $\mathcal{D}( p(t) \| \pref{}(\pi_{t}) )\ge 0$.  

\subsection{A Gauge Theory Perspective}
\label{sec:gauge}

We now show that the decomposition in Eq.~\eqref{eq:eprMaster} can be interpreted as the consequence of the gauge freedom discussed by Polettini in Ref.~\cite{polettini12}.
Indeed, in this reference he shows that the following gauge transformation leaves the stochastic dynamics \eqref{eq:ME} and the EP rate \eqref{eq:epr} unchanged:
\begin{equation}
	\begin{aligned}
		p_{n} & \rightarrow p_{n} \exp \gauge{n} \, , &
		w_{e} & \rightarrow w_{e} \exp - \gauge{\o{e}} \, , \\
		D^{n}_{e} & \rightarrow D^{n}_{e} \exp \gauge{n} \, , \; \text{and} &
		{\textstyle\sum_{n}} & \rightarrow {\textstyle\sum_{n}} \exp - \gauge{n} \, .
	\end{aligned}
	\label{eq:gauge}
\end{equation}
When considering a gauge term $\gauge{n}$ changing in time, one needs also to shift the time derivative as:
\begin{equation}
	\dt \rightarrow \dt - \partial_{t} \, ,
	\label{eq:gaugeDt}
\end{equation}
where $\partial_{t}$ behaves as a normal time derivative but it acts only on $\gauge{n}$.
Let us now consider the EP rate rewritten as
\begin{equation}
	\epr = \avef{j^{e}} \ln \frac{w_{e}}{w_{-e}} + \dt {\textstyle\sum_{n}} p_{n} \left[ - \ln p_{n} \right] \, .
	\label{}
\end{equation}
One readily sees that the transformation\eqref{eq:gauge}--\eqref{eq:gaugeDt} changes the first term into the nonconservative term, Eq.~\eqref{eq:eprMasterNC}, whereas the second into the conservative one, Eq.~\eqref{eq:eprMasterC}. 
We finally note that connections between gauge transformations and FTs were also discussed in Refs.~\cite{chetrite11,garrahan16}.

\bigskip

This concludes the presentation of our main results. 
In the following, we will consider various specific choices for $\pref{n}$ which solely depend on the driving protocol and thus give rise to DFTs. 
Each of them will provide a specific meaning to $\vep{nc}$ and $\vep{c}$.
Tab.~\ref{tab:master} summarizes the reference potential, affinity, and conservative contribution for these different choices.

\newcolumntype{L}{>{$\displaystyle}l<{$}}
\newcolumntype{C}{>{$\displaystyle}c<{$}}
\newcolumntype{R}{>{$\displaystyle}r<{$}}

\begin{table*}[t]
	\centering
	\begin{tabular}{R@{\quad}C@{\quad}C@{\quad}C@{\quad}C}
		\toprule
		\textbf{Decomposition}					&\bm{\Pref{n}}																&\bm{\Aref{}} 																											&\bm{\vepr{c}} 																				\\
		\midrule
		\text{adiabatic--nonadiabatic}			& - \ln \p{ss}{n} 															& \ln \frac{w_{e} \pss{\o{e}}}{w_{-e} \pss{\o{-e}}} 																& - \avef{j^{e}} D^{n}_{e} \ln \left\{ p_{n} / \p{ss}{n} \right\} 							\\[2.4ex]
		\text{cycle--cocycle}					& - \ln \left\{ {\textstyle\prod_{e \in \ST_{n}}} w_{e} - Z \right\} 		& \begin{cases} 0 \, , & \text{if } e \in \ST \, , \\ \mathcal{A}_{e} \, , & \text{if } e \in \chords \end{cases} 	& {\textstyle\sum_{e \in \ST}} \avef{\mathcal{J}_{e}} \, A_{e} 								\\[2.4ex]
		\text{system--reservoir}				& \mathcal{S}_{\mathrm{mc}} - S_{n}											& \delta S^{\mathrm{r}}_{e} = - f_{y} \deltaX^{y}_{e}																& \left[ S_{n} - \ln p_{n} \right] D^{n}_{e} \avef{j^{e}}									\\[1ex]
		\text{conservative--nonconservative}	& \Phi_{\mathrm{gg}} - \left[ S_{n} - F_{\lambda} L^{\lambda}_{n} \right]	& \mathcal{F}_{\yf} \deltaX^{\yf}_{e} 																				& \left[ S_{n} - F_{\lambda} L^{\lambda}_{n} - \ln p_{n} \right] D^{n}_{e} \avef{j^{e}} 	\\
		\bottomrule
	\end{tabular}
	\caption{
		Summary of the reference potentials, affinities, and conservative EP contributions for the specific references discussed in the text.
		The nonconservative EP contribution follows from $\vepr{nc} = \Aref{} \avef{j^{e}}$, whereas the driving one from $\vepr{d} = {\textstyle\sum_{n}} p_{n} \dt \Pref{n}$.
		Overall, $\epr = \vepr{nc} + \vepr{c} = \vepr{nc} + \vepr{d} - \dt \mathcal{D}(p\|\pref{})$, where $\mathcal{D}$ is the relative entropy.
	}
	\label{tab:master}
\end{table*}

\section{Adiabatic--Nonadiabatic Decomposition}
\label{sec:sse}

We now provide a first instance of reference PMF based on the fixed point of the Markov jump process.

The Perron--Frobenius theorem ensures that the ME \eqref{eq:ME} has, at all times, a unique instantaneous \emph{steady-state PMF} 
\begin{equation}
	{\textstyle\sum_{m}} W_{nm}(\pi_{t}) \pss{m}(\pi_{t}) = 0 \, , \quad \forall n \text{ and } t \, .
	\label{eq:prefSS}
\end{equation}

When using this PMF as the reference, $\pref{n}$ = $\pss{n}$, we recover the \emph{adiabatic--nonadiabatic EP rate} decomposition \cite{esposito07,esposito10:threedft,esposito10:threefaces1,ge10,garcia-garcia10,garcia-garcia12}.
More specifically, the nonconservative term gives the \emph{adiabatic} contribution which is zero only if the steady state satisfies detailed balance, and the conservative term gives the \emph{nonadiabatic} contribution which characterizes transient and driving effects.
A specific feature of this decomposition is that both terms are non-negative, as proved in App.~\ref{sec:ftMasterProofAdNonad}: $\vepr{nc} \geq 0$ and $\vepr{c} \geq 0$.
In turn, the nonadiabatic contribution decomposes into a relative entropy term and a driving one.

Provided that the forward and backward processes start in the steady state corresponding to the initial value of the respective protocol, the general DFT and IFT derived in Eqs.~\eqref{eq:dftMaster} and \eqref{eq:iftMaster} hold for the adiabatic and driving contributions of the adiabatic--nonadiabatic EP decomposition \cite{esposito07,esposito10:threedft}.

In detailed-balanced systems, the adiabatic contribution is vanishing, $\vepr{a} = 0$, and we obtain a FT for the sole driving contribution:
\begin{equation}
	\frac{P_{t}(\vep{d})}{P_{t}^{\dagger}(-\vep{d})} = \exp \vep{d} \, .
	\label{eq:dftSSdb}
\end{equation}
The celebrated Crooks' DFT \cite{crooks98,crooks99,crooks00} and Jarzynski's IFT \cite{jarzynski97:free} are of this type. 

\subsection*{Additional FTs}

Due to the particular mathematical properties of the steady-state PMF, additional FTs for the adiabatic and driving terms ensue.
These are not covered by our main DFT, Eq.~\eqref{eq:dftMaster}, and their proofs are discussed in App.~\ref{sec:dftAdNonadProof}.

For the former, the forward process is produced by the original dynamics initially prepared in an arbitrary PMF.
The backward process instead has the same initial PMF and the same driving protocol as the forward process, but the dynamics is governed by the rates
\begin{equation}
	\hat{w}_{e} := w_{-e} \pss{\o{-e}} / \pss{\o{e}} \, .
	\label{eq:fictRates}
\end{equation}

At any time, the following DFT relates the two processes,
\begin{equation}
	\frac{P_{t}(\vep{a})}{\hat{P}_{t}(-\vep{a})} = \exp \vep{a} \, ,
	\label{eq:DFTa}
\end{equation}
where ${\hat{P}~(-\vep{a})}$ is the probability of observing $-\vep{a}$ adiabatic EP during the backward process.
The Speck--Seifert IFT for the housekeeping heat is the IFT version of this DFT \cite{speck05}.

For the driving term, the forward process is again produced by the original dynamics, but now initially prepared in a steady state. 
The backward process is instead produced by the rates \eqref{eq:fictRates} with time-reversed driving protocol and the system must initially be prepared in a steady state.
Under these conditions, one has
\begin{equation}
	\frac{P_{t}(\vep{d})}{\hat{P}_{t}^{\dagger}(-\vep{d})} = \exp \vep{d} \, ,
	\label{eq:DFTd}
\end{equation}
where ${\hat{P}^{\dagger}(-\vep{d})}$ is the probability of observing $-\vep{d}$ driving EP during the backward process.
The Hatano--Sasa IFT \cite{hatano01} is the IFT version of this DFT.

\section{Cycle--Cocycle Decomposition}
\label{sec:ste}

We proceed by providing a second instance of reference PMF based on the equilibrium PMF for a spanning tree of the graph defined by the incidence matrix of the Markov jump process.

We partition the edges of the graph into two disjoint subsets: $\ST$ and $\chords$. 
The former identifies a \emph{spanning tree}, namely a minimal subset of paired edges, $(e,-e)$, that connects all states.
These edges are called \emph{cochords}.
All the other edges form $\chords$, and are called \emph{chords}.
Equivalently, $\ST$ is a maximal subset of edges that does not enclose any cycle---the trivial loops composed by forward and backward transitions, $(e,-e)$, are not regarded as cycles.
The graph obtained by combining $\ST$ and $e \in \chords$ identifies one and only one cycle, denoted by $\cycle_{e}$, for $e \in \chords$.
Algebraically, cycles are characterized as:
\begin{equation}
	\sum_{e' \in \cycle_{e}} D^{n}_{e'}
	= \sum_{e'} D^{n}_{e'} \, \cycle^{e'}_{e}
	= 0 \, , \quad \forall \, n \, ,
	\label{eq:nulright}
\end{equation}
where $\set{\cycle^{e'}_{e}}$, for $e \in \chords$, represent the vectors in the edge space whose entries are all zero except for those corresponding to the edges of the cycle, which are equal to one.

We now note that if $\ST$ were the sole allowed transitions, the PMF defined as follows would be an equilibrium steady state \cite{schnakenberg76}:
\begin{equation}
	\p{st}{n}(\pi_{t}) := \frac{1}{Z} \prod_{e \in \ST_{n}} w_{e}(\pi_{t}) \, ,
	\label{eq:prefST}
\end{equation}
where $Z = {\sum_{m}} \prod_{e \in \ST_{m}} w_{e}$ is a normalization factor, and $\ST_{n}$ denotes the spanning tree \emph{rooted} in $n$, namely the set of edges of $\ST$ that are oriented towards the state $n$.
Indeed, $\p{st}{n}$ would satisfy the property of detailed balance, Eq.~\eqref{eq:db}:
\begin{equation}
	\begin{split}
		w_{e} \p{st}{\o{e}}
		& = \frac{w_{e}}{Z} \prod_{e' \in \ST_{\o{e}}} w_{e'} = \frac{w_{-e}}{Z} \prod_{e' \in \ST_{\o{-e}}} w_{e'} \\
		& = w_{-e} \p{st}{\o{-e}}
		\, , \quad \forall \, e \in \ST \, .
	\end{split}
	\label{}
\end{equation}

We now pick this equilibrium PMF as a reference for our EP decomposition, $\pref{n}=\p{st}{n}$. 
However, in order to derive the specific expressions for $\vepr{nc}$ and $\vepr{c}$, the following result is necessary:
the edge probability fluxes can be decomposed as
\begin{equation}
	\avef{j^{e}} = {\sum_{e' \in \ST}} \avef{\mathcal{J}_{e'}} \edge^{e}_{e'} + {\sum_{e' \in \chords}} \avef{\mathcal{J}_{e'}} \cycle^{e}_{e'} \, ,
	\label{eq:JcyedDec}
\end{equation}
where $\set{\edge_{e}}$ denotes the canonical basis of the edge vector space: $\edge^{e'}_{e} = \delta^{e'}_{e}$ \cite{knauer11}.
Algebraically, this decomposition hinges on the fact that the set $\set{\cycle_{e}}_{e \in \chords} \cup \set{\edge_{e}}_{e \in \ST}$ is a basis of the edge vector space \cite{polettini14:cocycle}.
Note that for $e \in \chords$, the only nonvanishing contribution in Eq.~\eqref{eq:JcyedDec} comes from the cycle identified by $e$, and hence $\avef{j^{e}} = \avef{\mathcal{J}_{e}}$.
The coefficients $\set{\avef{\mathcal{J}_{e}}}$ are called \emph{cocycle fluxes} for the cochords, $e \in \ST$, and \emph{cycle fluxes} for the chords, $e \in \chords$.
They can be understood as follows \cite{polettini14:cocycle}:
removing a pair of edges, $e$ and $-e$, from the spanning tree ($e,-e \in \ST$) disconnects two blocks of states.
The cocycle flux $\set{\avef{\mathcal{J}_{e}}}$ of that edge is the probability flowing from the block identified by the origin of $e$, $\o{e}$, to that identified by the target of $e$, $\t{e}$.
Instead, the cycle flux $\set{\avef{\mathcal{J}_{e}}}$ of an edge, $e \in \chords$, quantifies the probability flowing along the cycle formed by adding that edge to the spanning tree.
Graphical illustrations of cocycle and cycle \emph{currents}, $\avef{\mathcal{J}^{e}} - \avef{\mathcal{J}^{-e}}$, can be found in Ref.~\cite{polettini14:cocycle}.

We can now proceed with our main task.
Using Eqs.~\eqref{eq:nulright} and \eqref{eq:prefST}, we verify that
\begin{equation}
	\Pref{n} D^{n}_{e} =
	\begin{cases}
		- \ln \left\{ w_{e}/w_{-e} \right\} \,,  						& \text{if } e \in \ST \, , \\
		- \ln \left\{ w_{e}/w_{-e} \right\} + \mathcal{A}_{e} \, ,	& \text{if } e \in \chords \, ,
	\end{cases}
	\label{eq:DlnProdw}
\end{equation}
where
\begin{equation}
	\mathcal{A}_{e} = {\textstyle\sum_{e'}} \cycle^{e'}_{e} \ln \left\{ {w_{e'}}/{w_{-e'}} \right\} \, , \quad \text{for } e \in \chords \, ,
	\label{}
\end{equation}
is the cycle affinity related to $\cycle_{e}$.
It follows that 
\begin{equation}
	\begin{split}
		\Aref{} & = \ln \frac{w_{e}}{w_{-e}} + \Pref{n} D^{n}_{e} = 
		\begin{cases}
			0 \, , 							& \text{if } e \in \ST \, , \\
			\mathcal{A}_{e} \, , 		 	& \text{if } e \in \chords \, ,
		\end{cases}
	\end{split}
	\label{Arefcycle}
\end{equation}
from which the nonconservative contribution readily follows:
\begin{equation}
	\vepr{nc} = {\sum_{e \in \chords}} \mathcal{A}_{e} \avef{j_{e}} = {\sum_{e \in \chords}} \mathcal{A}_{e} \avef{\mathcal{J}_{e}} \, .
	\label{eq:eprSTnc}
\end{equation}
In the last equality, we used the property of cycle fluxes discussed after Eq.~\eqref{eq:JcyedDec}.
Hence, the nonconservative contribution accounts for the dissipation along network cycles.
In turn, combining Eq.~\eqref{eq:eprMasterC} with Eqs.~\eqref{eq:JcyedDec} and \eqref{eq:DlnProdw}, one obtains the conservative contribution
\begin{equation}
	\vepr{c} = {\sum_{e \in \ST}} \, A_{e} \avef{\mathcal{J}_{e}} \, ,
	\label{eq:eprSTc}
\end{equation}
which accounts for the dissipation along cocycles.
Using these last two results, the EP decomposition \eqref{eq:eprMaster} becomes the \emph{cycle--cocycle} decomposition found in Ref.~\cite{polettini14:cocycle}:
\begin{equation}
	\epr = {\sum_{e \in \chords}} \mathcal{A}_{e} \avef{j_{e}} + {\sum_{e \in \ST}} \, A_{e} \avef{\mathcal{J}_{e}} \, .
	\label{eq:eprST}
\end{equation}

As for all decompositions, the conservative contribution---here the cocycle one---vanishes at steady state in the absence of driving.
The cycle contribution instead disappears in detailed-balanced systems, when all the cycle affinities vanish.
This statement is indeed the \emph{Kolmogorov criterion} for detailed balance \cite{kolmogoroff36,kelly79}.

The fluxes decomposition Eq.~\eqref{eq:JcyedDec} is also valid at the trajectory level, where the cycle and cocycle fluxes become fluctuating instantaneous fluxes, $\set{\mathcal{J}_{e}}$. 
Obviously, the same holds true for the cycle--cocycle EP decomposition.
Therefore, if the system is in an equilibrium PMF of type \eqref{eq:prefST} at the beginning of the forward and the backward process, a DFT and an IFT hold by applying Eqs.~\eqref{eq:dftMaster} and \eqref{eq:iftMaster}.
Note that the fluctuating quantity appearing in the DFT, $\vep{d} + \vep{nc}$, can be interpreted as the EP of the extended process in which, at time $t$, the driving is stopped, all transitions in $\chords$ are shut down, and the system is allowed to relax to equilibrium---which is the initial PMF of the backward process.

It is worth mentioning that one can easily extend the formulation of our DFT by considering the joint probability distribution for each subcontribution of $\vep{d}$ and $\vep{na}$ antisymmetrical under time reversal.
This can be shown using either the proof in App.~\ref{sec:ftMasterProof2} \cite{garcia-garcia10}, or that in App.~\ref{sec:ftMasterProof} \cite{rao18:shape}.
In the case of the cycle--cocycle decomposition, it would lead to
\begin{multline}
	\frac{P_{t}(\vep{d},\set{ \mathcal{A}_{e} \left( j_{e} - j_{-e} \right)}_{e \in \chords})}{P_{t}^{\dagger}(-\vep{d},\set{ - \mathcal{A}_{e} \left( j_{e} - j_{-e} \right)}_{e \in \chords})} \\
	= \exp\left\{ \vep{d} + {\sum_{e \in \chords}} \mathcal{A}_{e} j_{e} \right\} \, ,
	\label{eq:dftCycleCurrents}
\end{multline}
which is a generalization of the DFT derived in Ref.~\cite{polettini14:transient} to time-inhomogeneous systems.
In turn, the latter is a generalization of the steady-state DFT derived by Andrieux and Gaspard in Ref.~\cite{andrieux07:schnakenberg} to finite times.

\begin{figure}[tb]
	\centering
	\subfloat[][Spanning Tree]
	{\includegraphics[width=.25\textwidth]{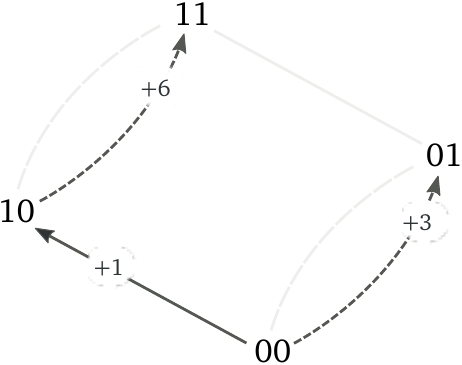} \label{fig:st} } \quad \quad \quad
	\subfloat[][Cycles]
	{\includegraphics[width=.48\textwidth]{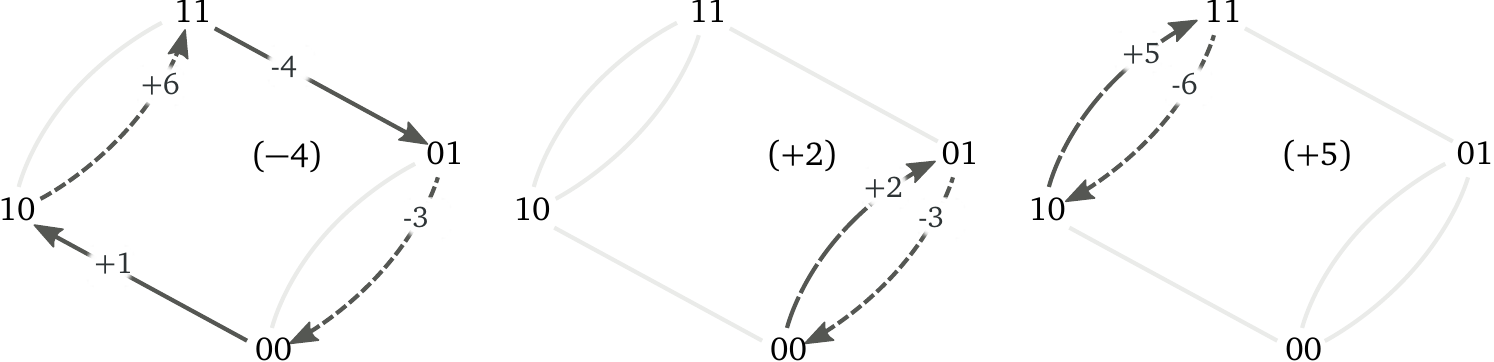} \label{fig:cycles} }
	\caption{
		(\textbf{a}) Spanning tree, and (\textbf{b}) corresponding cycles for the network in Fig.~\ref{fig:network}.
	}
	\label{fig:cycle--cocycle}
\end{figure}

\paragraph*{Example}
A spanning tree for the network in Fig.~\ref{fig:network} is depicted in Fig.~\ref{fig:st}.
The cycles defined by the corresponding chords are depicted in Fig.~\ref{fig:cycles}.
Algebraically, these cycles are represented as
\begin{equation}
	\mathcal{C} =
	\kbordermatrix{
		& \greyt{-4} & \greyt{+2} & \greyt{+5} \\
		\greyt{+1} & 1 & 0 & 0 \\
		\greyt{+2} & 0 & 1 & 0 \\
		\greyt{+3} & -1 & -1 & 0 \\
		\greyt{+4} & -1 & 0 & 0 \\
		\greyt{+5} & 0 & 0 & 1 \\
		\greyt{+6} & 1 & 0 & -1
	} \, ,
	\label{eq:exCycles}
\end{equation}
where the negative entries must be regarded as transitions performed in the backward direction.
The corresponding affinities, which determine the nonconservative contribution \eqref{eq:eprSTnc}, hence read:
\begin{equation}
	\begin{aligned}
		\mathcal{A}_{+2} & = \ln \frac{w_{+2}w_{-3}}{w_{-2}w_{+3}} \, , \quad
		\mathcal{A}_{+5} = \ln \frac{w_{+5}w_{-6}}{w_{-5}w_{+6}} \, , \; \text{and} \\
		\mathcal{A}_{-4} & = \ln \frac{w_{+1}w_{+6}w_{-4}w_{-3}}{w_{-1}w_{-6}w_{+4}w_{+3}} \, .
	\end{aligned}
	\label{}
\end{equation}
The affinities corresponding to the cycles taken in the backward direction follow from $\mathcal{A}_{-e} = - \mathcal{A}_{e}$.
Regarding the expression of the cocycle fluxes, it can be checked that they are equal to
\begin{equation}
	\hspace{-2.4em} 
	\begin{aligned}
		\avef{\mathcal{J}_{+1}} & = \avef{j_{+1}} - \avef{j_{-4}} \, , &
		\avef{\mathcal{J}_{-1}} & = \avef{j_{-1}} - \avef{j_{+4}} \, , \\
		\avef{\mathcal{J}_{+3}} & = \avef{j_{+3}} - \avef{j_{-2}} - \avef{j_{+4}} \, , &
		\avef{\mathcal{J}_{-3}} & = \avef{j_{-3}} - \avef{j_{+2}} - \avef{j_{-4}} \, , \\
		\avef{\mathcal{J}_{+6}} & = \avef{j_{+6}} - \avef{j_{-5}} - \avef{j_{-4}} \, , &
		\avef{\mathcal{J}_{-6}} & = \avef{j_{-6}} - \avef{j_{+5}} - \avef{j_{+4}} \, ,
	\end{aligned}
	\label{}
\end{equation}
by expanding Eq.~\eqref{eq:eprST} into Eq.~\eqref{eq:epr}.

\section{Stochastic Thermodynamics}
\label{stochTherm}

The results obtained until this point are mathematical and have a priori no connection to physics.
We now specify the conditions under which a Markov jump process describes the dynamics of an open physical system in contact with multiple reservoirs.
This will enable us to introduce physically motivated decompositions and derive DFTs with a clear thermodynamic interpretation.
 
\begin{figure}[tb]
	\centering
	\includegraphics[width=.40\textwidth]{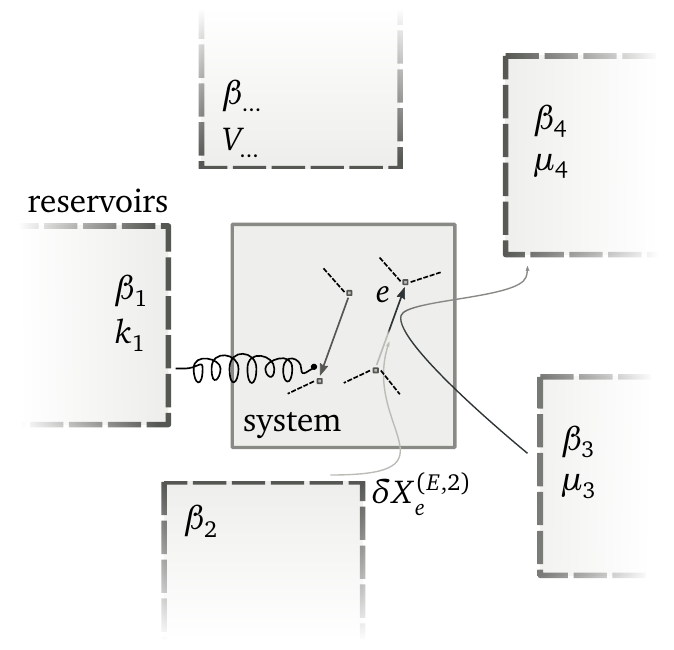}
	\caption{\label{fig:systemBath}
		Pictorial representation of a system coupled to several reservoirs.
		Transitions may involve more than one reservoir and exchange between reservoirs.
		Work reservoirs are also taken into account.
	}
\end{figure}

Each system state, $n$, is now characterized by given values of some \emph{system quantities}, $\set{\X_{n}}$, for $\lX=1,~\dots,~\nof{\kappa}$, which include the internal energy, $E_{n}$, and possibly additional ones (see Tab.~\ref{tab:fYpairs} for some examples).
These must be regarded as globally \emph{conserved quantities}, as their change in the system is always balanced by an opposite change in the reservoirs.
When labeling the reservoirs with $\set{r}$, for $r=1,\dots,\nof{\mathrm{r}}$, the \emph{balance equation} for $\X$ along the transition $e$ can be written as:
\begin{equation}
	{\X_{n'} D^{n'}_{e}} = \delta_{\mathrm{i}} \X_{e} + {\textstyle\sum_{r}} {\deltaX^{(\lX,r)}_{e}} \, .
	\label{eq:balanceGlobal}
\end{equation}
The lhs is the overall change in the system, whereas $\delta_{\mathrm{i}} \X_{e}$ denotes the changes due to internal transformations (\emph{e.g.}, chemical reactions \cite{schmiedl07,rao18}), and $\deltaX^{(\lX,r)}_{e}$ quantifies the amount of $\X$ supplied by the reservoir $r$ to the system along the transition $e$.
For the purposes of our discussion, we introduce the index $y=(\lX,r)$---\emph{i.e.}, \emph{the conserved quantity $\X$ exchanged with the reservoir} $r$---and define the matrix $\deltaX$ whose entries are $\set{\deltaX^{y}_{e} \equiv \deltaX^{(\lX,r)}_{e}}$.
All indices used in the following discussion are summarized in Tab.~\ref{tab:indices}.
Microscopic reversibility requires that $\deltaX^{y}_{-e} = - \deltaX^{y}_{e}$.
Note that more than one reservoir may be involved in each transition (see Fig.~\ref{fig:systemBath}).

\begin{table}[tb]
	\centering
	\begin{tabular}{lr}
		\toprule
		\textbf{System Quantity \bm{$\X${  }}} 	& \textbf{Intensive Field \bm{$f_{(\lX,r)}$}}\\
		\midrule
		energy, $E_{n}$					& inverse temperature, $\beta_{r}$	\\
		particles number, $N_{n}${ }	& chemical potential, $- \beta_{r} \mu_{r}$ \\
		charge, $Q_{n}$					& electric potential, $- \beta_{r} V_{r}$	\\
		displacement, $X_{n}$			& generic force, $- \beta_{r} k_{r}$ \\
		angle, $\theta_{n}$				& torque, $- \beta_{r} \tau_{r}$ \\
		\bottomrule
	\end{tabular}
	\caption{
		Examples of system quantity--intensive field conjugated pairs in the entropy representation.
		$\beta_{r} := 1/T_{r}$ denotes the inverse temperature of the reservoir.
		Since charges are carried by particles, the conjugated pair $(Q_{n}, - \beta_{r} V_{r})$ is usually embedded in $(N_{n}, - \beta_{r} \mu_{r})$.
	}
	\label{tab:fYpairs}
\end{table}

\begin{table}[tb]
	\centering
	\begin{tabular}{lcr}
		\toprule
		\textbf{Index{ }}	& \textbf{Label for{  }} 	& \textbf{Number} \\
		\midrule
		$n$			& state 					& $\nof{\mathrm{n}}$ \\
		$e$			& transition 				& $\nof{\mathrm{e}}$ \\
		$\lX$ 		& system quantity 			& $\nof{\kappa}$ \\
		$r$ 		& reservoir		 			& $\nof{\mathrm{r}}$ \\
		$y \equiv (\lX,r)${ }			& conserved quantity $\X$ from reservoir $r$		& $\nof{\mathrm{y}}$ \\
		$\lambda$	& conservation law and conserved quantity 	& $\nof{\lambda}$ \\
		$\yp$		& ``potential'' $y$ 	& $\nof{\lambda}$ \\
		$\yf$		& ``force'' $y$		& $\nof{\mathrm{y}} - \nof{\lambda}$ \\
		\bottomrule
	\end{tabular}
	\caption{
		Summary of the indices used throughout the paper and the object they label.
	}
	\label{tab:indices}
\end{table}

In addition to the trivial set of conserved quantities $\set{\X}$, the system may be characterized by some additional ones, which are \emph{specific} for each system.
We now sketch the systematic procedure to identify these quantities and the corresponding conservation laws \cite{polettini16,rao18:shape}.
Algebraically, conservation laws can be identified as a maximal set of independent vectors in the $y$-space, $\set{\bm{\ell}^{\lambda}}$, for $\lambda = 1,~\dots,~\nof{\lambda}$, such that
\begin{equation}
	\hspace{-1em}
	\ell^{\lambda}_{y} \, \deltaX^{y}_{e'} \, \cycle^{e'}_{e} = 0 \, , \quad \text{for all cycles, \emph{i.e.}, for all } e \in \chords \, .
	\label{eq:conservationLaws}
\end{equation}
Indeed, the quantities $\set{\ell^{\lambda}_{y} \, \deltaX^{y}_{e}}$, for $\lambda = 1,~\dots,~\nof{\lambda}$, are combinations of exchange contributions $\set{\deltaX^{y}_{e}}$, for $y=1,~\dots,~\nof{\lambda}$, which vanish along all cycles.
They must therefore identify some state variables, $\set{L^{\lambda}}$, for $\lambda = 1,~\dots,~\nof{\mathrm{y}}$, in the same way curl-free vector fields are conservative and identify scalar potentials:
\begin{equation}
	L^{\lambda}_{n} \, D^{n}_{e} = \ell^{\lambda}_{y} \, \deltaX^{y}_{e} \equiv {\textstyle\sum_{r}} \left\{ {\textstyle\sum_{\lX}}\ell^{\lambda}_{(\lX,r)} \, \deltaX^{(\lX,r)}_{e} \right\} \, .
	\label{eq:components}
\end{equation}
This equation can be regarded as the balance equation for the conserved quantities.
In the absence of internal transformations, $\delta_{\mathrm{i}} \X_{e}$, trivial conservation laws correspond to $\ell^{\lX}_{y} \equiv \ell^{\lX}_{(\lX',r)} = \delta^{\lX}_{\lX'}$, so that the balance Eqs.~\eqref{eq:balanceGlobal} are recovered.
Notice that each $L^{\lambda}$ is defined up to a reference value.

Each reservoir $r$ is characterized by a set of \emph{entropic intensive fields} conjugated to the exchange of the system quantities $\set{\X}$, $\set{f_{(\lX,r)}}$ for ${\lX=1,~\dots,~\nof{\kappa}}$ (\emph{e.g.},~\cite{callen85} \S~2-3).
A short list of $\X$--$f_{(\lX,r)}$ conjugated pairs is reported in Tab.~\ref{tab:fYpairs}.
The thermodynamic consistency of the stochastic dynamics is ensured by the \emph{local detailed balance},
\begin{equation}
	\ln \frac{w_{e}}{w_{-e}} = - f_{y} \deltaX^{y}_{e} + S_{n} D^{n}_{e} \, .
	\label{eq:ldb}
\end{equation}
It relates the log ratio of the forward and backward transition rates to the entropy change in the reservoirs resulting from the transfer of system quantities during that transition.
This entropy change is evaluated using equilibrium thermodynamics (in the reservoirs), and reads $\set{\delta S^{\mathrm{r}}_{e} = - f_{y} \deltaX^{y}_{e}}$.
The second term on the rhs is the internal entropy change occurring during the transition, as $S_{n}$ quantifies the internal entropy of the state $n$.
This term can be seen as the outcome of a coarse-graining procedure over a finer description in which multiple states with the same system quantities are collected in one single $n$ \cite{esposito12}.
Using Eq.~\eqref{eq:ldb}, the affinities, Eq.~\eqref{eq:affinity}, can be rewritten as:
\begin{equation}
	A_{e} = {\textstyle\sum_{r}} \left[ - {\textstyle\sum_{\lX}} f_{(\lX,r)} \deltaX^{(\lX,r)}_{e} \right] + \left[ S_{n} - \ln p_{n} \right] D^{n}_{e} \, .
	\label{eq:entBalanceEdge}
\end{equation}
This relation shows that the affinity is the entropy change in all reservoirs plus the system entropy change.
In other words, while Eq.~\eqref{eq:components} characterizes the balance of the conserved quantities along the transitions, Eq.~\eqref{eq:entBalanceEdge} characterizes the corresponding lack of balance for entropy, namely the second law.

As for the transition rates, the changes in time of the internal entropy $S$, the conserved quantities $\set{\X}$ (hence $\set{\deltaX^{y}_{e}}$), and their conjugated fields $\set{f_{y}}$, are all encoded in the protocol function $\pi_{t}$.
Physically, this modeling describes the two possible ways of controlling a system:
either through $\set{\X}$ or $S$ which characterize the system states, or through $\set{f_{y}}$ which characterize the properties of the reservoirs.

\begin{figure*}[t]
	\centering
	\subfloat[][Scheme]
	{\includegraphics[width=.35\textwidth]{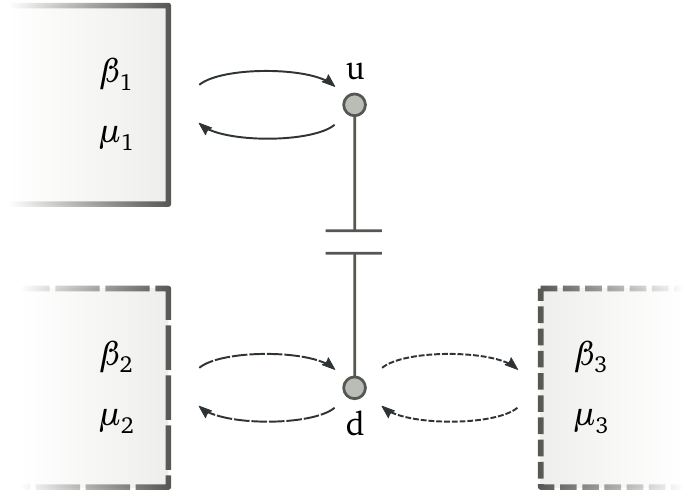} \label{fig:qd} } \quad\quad\quad\quad
	\subfloat[][Energy Landscape]
	{\includegraphics[width=.35\textwidth]{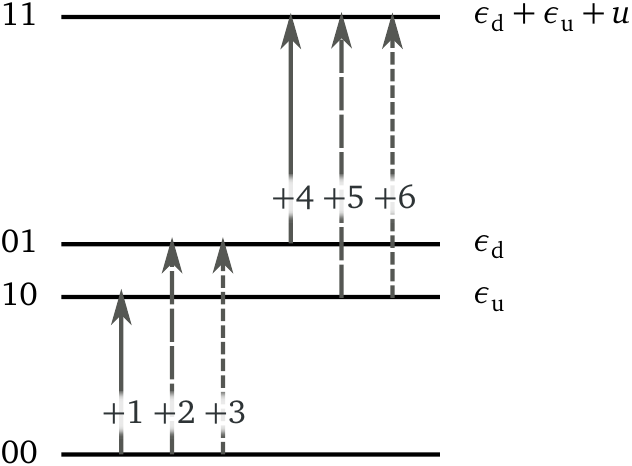} \label{fig:energy} }
	\caption{
		Double coupled quantum dot (QD) in contact with three reservoirs.
		Transitions related to the first reservoir are depicted using solid lines, while those related to the second and third ones using dashed and dotted lines, respectively.
		The graphical rule was applied to the network of transitions in Fig.~\ref{fig:network}.
		(\textbf{a}) Pictorial representation of the system.
		The upper dot $\mathrm{u}$ is in contact with the first reservoir, while the lower dot $\mathrm{d}$ with the second and third reservoirs.
		Energy and electrons are exchanged, but the dots cannot host more than one electron.
		(\textbf{b}) Energy landscape of the dot.
		When both dots are occupied, $11$, a repulsive energy $u$ adds to the occupied dots energies, $\epsilon_{\mathrm{u}}$ and $\epsilon_{\mathrm{d}}$.
	}
	\label{fig:scheme}
\end{figure*}

\paragraph*{Example}
We illustrate the role of system-specific conservation laws by considering the double quantum dot (QD) depicted in Fig.~\ref{fig:qd} \cite{sanchez12,strasberg13,thierschmann15}, whose network of transition and energy landscape are drawn in Figs.~\ref{fig:network} and \ref{fig:energy}, respectively.
Electrons can enter empty dots from the reservoirs, but cannot jump from one dot to the other.
When the two dots are occupied, an interaction energy, $u$, arises.
Energy, $E_{n}$, and total number of electrons, $N_{n}$, characterize each state of the system:
\begin{equation}
	\hspace{-2.6em}
	\begin{aligned}
		E_{00} & = 0 \, , & E_{10} & = \epsilon_{\mathrm{u}} \, , 	& E_{01} & = \epsilon_{\mathrm{d}} \, , & E_{11} & = \epsilon_{\mathrm{u}} + \epsilon_{\mathrm{d}} + u , \\
		N_{00} & = 0 \, , & N_{10} & = 1 \, , 						& N_{01} & = 1 \, , & N_{11} & = 2 \, ,
	\end{aligned}
	\label{eq:cqQD}
\end{equation}
where the first entry in $n$ refers to the occupancy of the upper dot, and the second to the lower.
\\The entries of the matrix $\deltaX$ for the forward transitions are:
\begin{equation}
	\hspace{-1em}
	\deltaX =
	\kbordermatrix{
		& \greyt{+1} & \greyt{+2} & \greyt{+3} & \greyt{+4} & \greyt{+5} & \greyt{+6} \\
		\greyt{(E,1)} & \epsilon_{\mathrm{u}} & 0 & 0 & \epsilon_{\mathrm{u}} + u & 0 & 0 \\
		\greyt{(N,1)} & 1 & 0 & 0 & 1 & 0 & 0 \\
		\greyt{(E,2)} & 0 & \epsilon_{\mathrm{d}} & 0 & 0 & \epsilon_{\mathrm{d}} + u & 0 \\
		\greyt{(N,2)} & 0 & 1 & 0 & 0 & 1 & 0 \\
		\greyt{(E,3)} & 0 & 0 & \epsilon_{\mathrm{d}} & 0 & 0 & \epsilon_{\mathrm{d}} + u \\
		\greyt{(N,3)} & 0 & 0 & 1 & 0 & 0 & 1
	} \
	\label{eq:exchangedQD}
\end{equation}
(see Fig.~\ref{fig:network}), whereas the entries related to backward transition follow from $\deltaX^{y}_{-e} = - \deltaX^{y}_{e}$.
For instance, along the first transition the system gains $\epsilon_{\mathrm{u}}$ energy and $1$ electron from the reservoir $1$.
The vector of entropic intensive fields is given by
\begin{equation}
	\hspace{-1em}
	\bm {f} =
	\kbordermatrix{
		& \greyt{(E,1)} & \greyt{(N,1)} & \greyt{(E,2)} & \greyt{(N,2)} & \greyt{(E,3)} & \greyt{(N,3)} \\
		& \beta_{1} & - \beta_{1} \mu_{1} & \beta_{2} & - \beta_{2} \mu_{2} & \beta_{3} & - \beta_{3} \mu_{3}
	} \, .
	\label{eq:intensiveQD}
\end{equation}

Since the QDs and the electrons have no internal entropy, $S_{n} = 0$ for all $n$, the local detailed balance property, Eq.~\eqref{eq:ldb}, can be easily recovered from the product $- \bm f \deltaX$.
From a stochastic dynamics perspective, this property arises when considering fermionic transition rates, namely $w_{e} = \Gamma_{e}(1+\exp\{ f_{y}\deltaX^{y}_{e} \})^{-1}$ and $w_{-e} = \Gamma_{e}\exp\{ f_{y}\deltaX^{y}_{e} \}(1+\exp\{ f_{y}\deltaX^{y}_{e} \})^{-1}$ for electrons entering and leaving the dot.

A maximal set of independent vectors in $y$-space satisfying Eq.~\eqref{eq:conservationLaws} is composed of
\begin{equation}
	\begin{aligned}
		\bm \ell^{\mathrm{E}} & =
		\kbordermatrix{
			& \greyt{(E,1)} & \greyt{(N,1)} & \greyt{(E,2)} & \greyt{(N,2)} & \greyt{(E,3)} & \greyt{(N,3)} \\
			& 1 & 0 & 1 & 0 & 1 & 0
		} \, , \\
		\bm \ell^{\mathrm{u}} & =
		\kbordermatrix{
			& \greyt{(E,1)} & \greyt{(N,1)} & \greyt{(E,2)} & \greyt{(N,2)} & \greyt{(E,3)} & \greyt{(N,3)} \\
			& 0 & 1 & 0 & 0 & 0 & 0
		} \, , \\
		\bm \ell^{\mathrm{d}} & =
		\kbordermatrix{
			& \greyt{(E,1)} & \greyt{(N,1)} & \greyt{(E,2)} & \greyt{(N,2)} & \greyt{(E,3)} & \greyt{(N,3)} \\
			& 0 & 0 & 0 & 1 & 0 & 1
		} \, .
	\end{aligned}
	\label{eq:allCLQD}
\end{equation}
The first vector identifies the energy state variable, $E_{n}$:
\begin{equation}
	\hspace{-2.2em}
	\bm \ell^{E} \deltaX = 
	\kbordermatrix{
		& \greyt{+1} & \greyt{+2} & \greyt{+3} & \greyt{+4} & \greyt{+5} & \greyt{+6} \\
		& \epsilon_{\mathrm{u}} & \epsilon_{\mathrm{d}} & \epsilon_{\mathrm{d}} & \epsilon_{\mathrm{u}} + u & \epsilon_{\mathrm{d}} + u & \epsilon_{\mathrm{d}} + u
	}
	\equiv \set{E_{n} D^{\mathrm{n}}_{e}} \, .
	\label{}
\end{equation}
The other two instead give the occupancy of the upper and lower dots, $N^{\mathrm{u}}_{n}$ and $N^{\mathrm{d}}_{n}$:
\begin{equation}
	\begin{aligned}
		\bm \ell^{\mathrm{u}} \deltaX & = 
		\kbordermatrix{
			& \greyt{+1} & \greyt{+2} & \greyt{+3} & \greyt{+4} & \greyt{+5} & \greyt{+6} \\
			& 1 & 0 & 0 & 1 & 0 & 0 \\
		}
		\equiv \set{N^{\mathrm{u}}_{n} D^{\mathrm{n}}_{e}} \, , \\
		\bm \ell^{\mathrm{d}} \deltaX & = 
		\kbordermatrix{
			& \greyt{+1} & \greyt{+2} & \greyt{+3} & \greyt{+4} & \greyt{+5} & \greyt{+6} \\
			& 0 & 1 & 1 & 0 & 1 & 1 \\
		}
		\equiv \set{N^{\mathrm{d}}_{n} D^{\mathrm{n}}_{e}} \, .
	\end{aligned}
	\label{}
\end{equation}
{A posteriori}, we see that these conservation laws arise from the fact that no electron transfer from one dot to the other is allowed.
The total occupancy of the system, $N_{n}$, is recovered from the sum of the last two vectors.

\bigskip

Now that a nonequilibrium thermodynamics has been built on top of the Markov jump process, we can proceed by considering two physical relevant $\pref{n}$.

\section{System--Reservoirs Decomposition}
\label{sec:mce}

We start by considering a microcanonical PMF as reference:  
\begin{equation}
	\pref{n} = p^{\mathrm{mc}}_{n} := {\exp\left\{ S_{n} - \mathcal{S}_{\mathrm{mc}} \right\}} \, ,
	\label{eq:prefMC}
\end{equation}
where
\begin{equation}
	\mathcal{S}_{\mathrm{mc}} = \ln {\textstyle\sum_{m}} \exp S_{m}
	\label{eq:boltzmann}
\end{equation}
is the \emph{Boltzmann's equilibrium entropy}.
With this choice, the reference affinities become sums of entropy changes in the reservoirs
\begin{equation}
	\Aref{} = \delta S^{\mathrm{r}}_{e} = - f_{y} \deltaX^{y}_{e} \, ,
	\label{}
\end{equation}
and hence the nonconservative contribution becomes the rate of entropy change in all reservoirs
\begin{equation}
	\vepr{nc} = \avef{\dot{S}_{\mathrm{r}}} = - f_{y} \deltaX^{y}_{e} \avef{j^{e}} \, .
	\label{eq:eprMCnc}
\end{equation}
For the conservative contribution, one instead obtains:
\begin{align}
	\vepr{c} = \left[ S_{n} - \ln p_{n} \right] D^{n}_{e} \avef{j^{e}} \, .
	\label{eq:veprMCc}
\end{align}
Using Eq.~\eqref{eq:eprMasterCD}, it can be rewritten in terms of the Gibbs--Shannon entropy,
\begin{equation}
	\avef{\mathcal{S}} = {\textstyle\sum_{n}} p_{n} \left[ S_{n} - \ln p_{n} \right]
	\label{eq:shannon}
\end{equation}
and the Boltzmann entropy.
Indeed,
\begin{equation}
	\mathcal{D}(p\|\p{mc}{}) = \mathcal{S}_{\mathrm{mc}} - \avef{\mathcal{S}}
	\label{}
\end{equation}
and
\begin{equation}
	\vepr{d} = \dt \mathcal{S}_{\mathrm{mc}} - {\textstyle\sum_{n}} p_{n} \dt S_{n} \, ,
	\label{eq:veprMCd}
\end{equation}
so that 
\begin{align}
	\vepr{c} = \dt \avef{\mathcal{S}} - {\textstyle\sum_{n}} p_{n} \dt S_{n} \, .
	\label{eq:eprMCcd}
\end{align}
The conservative contribution thus contains changes in the system entropy caused by the dynamics and the external drive. 

The EP decomposition \eqref{eq:eprMaster} with Eqs.~\eqref{eq:eprMCnc} and \eqref{eq:eprMCcd} is thus the well-known \emph{system--reservoir} decomposition (\emph{i.e.}, the traditional \emph{entropy balance}).
Since the same decomposition holds at the trajectory level, if the initial PMF of the forward and backward processes are microcanonical, the DFT and IFT hold by applying Eqs.~\eqref{eq:dftMaster} and \eqref{eq:iftMaster}.
When the driving does not affect the internal entropy of the system states $\set{S_{n}}$, the DFT and IFT hold for the reservoir entropy alone.
Finally, the fluctuating quantity appearing in the DFT, $\vep{d} + \vep{nc}$, can be interpreted as the EP of the extended process in which, at time $t$, the driving is stopped, all temperatures are raised to infinity, $\beta_{r}\rightarrow 0$, and the system is allowed to relax to equilibrium---the initial PMF of the backward process.

\section{Conservative--Nonconservative Decomposition}
\label{sec:gge}

We now turn to a reference PMF which accounts for conservation laws: the \emph{generalized Gibbs PMF}.

To characterize this PMFs, we observe that since $\set{\bm \ell^{\lambda}}$ are linearly independent (otherwise we would have linearly dependent conserved quantities), one can always identify a set of $y$'s, denoted by $\set{\yp}$, such that the matrix whose rows are $\set{\ell^{\lambda}_{\yp}}$, for $\lambda = 1,\dots,\nof{\lambda}$, is nonsingular.
We denote by $\set{\overline{\ell}^{\yp}_{\lambda}}$ for $\lambda = 1,\dots,\nof{\lambda}$, the columns of the inverse matrix.
All other $y$'s are denoted by $\set{\yf}$.
Using the splitting $\set{\yp}$--$\set{\yf}$ and the properties of $\set{\ell^{\lambda}_{\yp}}$, in combination with the balance equation for conserved quantities, Eq.~\eqref{eq:components}, the local detailed balance \eqref{eq:ldb} can be decomposed as
\begin{equation}
	\ln \frac{w_{e}}{w_{-e}} = \mathcal{F}_{\yf} \deltaX^{\yf}_{e} + \left[ S_{n} - F_{\lambda} L^{\lambda}_{n} \right] D^{n}_{e} \, ,
	\label{eq:ldbAwesome}
\end{equation}
where
\begin{equation}
	F_{\lambda} = f_{\yp} \overline{\ell}^{\yp}_{\lambda}
	\label{eq:effectiveIntensive}
\end{equation}
are the system-specific intensive fields conjugated to the conserved quantities, and
\begin{equation}
	\mathcal{F}_{\yf} := F_{\lambda} \, \ell^{\lambda}_{\yf} - f_{\yf}
	\label{eq:fundForce}
\end{equation}
are differences of intensive fields called nonconservative \emph{fundamental forces}.
Indeed, these nonconservative forces are responsible for breaking detailed balance. 
When they all vanish, $\mathcal{F}_{\yf} = 0$ for all $\yf$, the system is indeed detailed balanced and the PMF 
\begin{equation}
	p^{\mathrm{gg}}_{n} :=  {\exp \left\{ S_{n} - F_{\lambda} L^{\lambda}_{n} - \Phi_{\mathrm{gg}} \right\}} \, ,
	\label{eq:PrefGG}
\end{equation}
with $\Phi_{\mathrm{gg}} := \ln {\textstyle\sum_{n}} \exp\left\{ S_{n} - F_{\lambda} L^{\lambda}_{n} \right\}$, satisfies the detailed balance property \eqref{eq:db}.
The potential corresponding to Eq.~\eqref{eq:PrefGG}, $\P{gg}{n}$, is minus the \emph{Massieu potential} which is constructed by using all conservation laws (\emph{e.g.}~\cite{callen85} \S\S~5-4 and 19-1, \cite{peliti11} \S~3.13).
Choosing the PMF \eqref{eq:PrefGG} as a reference, $\pref{n} = p^{\mathrm{gg}}_{n}$,
the reference affinity straightforwardly ensues from Eq.~\eqref{eq:ldbAwesome},
\begin{equation}
	\Aref{} = \A{gg}{} = \mathcal{F}_{\yf} \deltaX^{\yf}_{e} \, .
	\label{eq:ArefGG}
\end{equation}
Hence,
\begin{equation}
	\vepr{nc} = \mathcal{F}_{\yf} \avef{I^{\yf}} \, ,
	\label{eq:eprGGnc}
\end{equation}
where
\begin{equation}
	\avef{I^{\yf}} = \deltaX^{\yf}_{e} \avef{j^{e}}
\end{equation}
are the fundamental currents conjugated to the forces.
For the conservative contribution, one obtains
\begin{align}
	\vepr{c} = \left[ S_{n} - F_{\lambda} L^{\lambda}_{n} - \ln p_{n} \right] D^{n}_{e} \avef{j^{e}} \, .
	\label{eq:eprGGc}
\end{align}
When written as in Eq.~\eqref{eq:eprMasterCD}, its two contributions are:
\begin{equation}
	\mathcal{D}(p\|\p{gg}{}) = \Phi_{\mathrm{gg}} - {\textstyle\sum_{n}} p_{n} \left[ S_{n} - F_{\lambda} L^{\lambda}_{n} - \ln p_{n} \right] \, ,
	\label{}
\end{equation}
which relates the equilibrium Massieu potential to its averaged nonequilibrium counterpart;
and
\begin{equation}
	\vepr{d} = \dt \Phi_{\mathrm{gg}} - {\textstyle\sum_{n}} p_{n} \dt \left[ S_{n} - F_{\lambda} L^{\lambda}_{n} - \ln p_{n} \right] \, ,
	\label{eq:veprGGd}
\end{equation}
which quantifies the dissipation due to external manipulations of $\set{S_{n}}$, the fields $\set{F_{\lambda}}$, and the conserved quantities $\set{L^{\lambda}}$. 
We emphasize that since $\P{gg}{n}$ encompasses all conserved quantities, $\vepr{c}$ captures all dissipative contributions due to conservative forces.
Hence, $\vepr{nc}$ consists of a minimal number, $\nof{\mathrm{y}} - \nof{\lambda}$, of purely nonconservative contributions.
The EP decomposition Eq.~\eqref{eq:eprMaster} with Eqs.~\eqref{eq:eprGGnc} and \eqref{eq:eprGGc} is the \emph{conservative--nonconservative} decomposition of the EP obtained in Ref.~\cite{rao18:shape}.

The conservative--nonconservative splitting of the EP can also be made at the trajectory level.
Hence, if the initial condition of the forward and backward process is of the form \eqref{eq:PrefGG}, the DFT and IFT given by Eqs.~\eqref{eq:dftMaster} and \eqref{eq:iftMaster} hold.
Here too, the fluctuating quantity appearing in the DFT, $\vep{d} + \vep{nc}$, can be interpreted as the EP of an extended process including relaxation, but for nonisothermal processes the procedure can be significantly more involved.
The details of this discussion can be found in Ref.~\cite{rao18:shape}.

\paragraph*{Example}
We now provide the expressions of $\Pref{n}$ and $\Aref{}$ for the double QD discussed in the previous example, Fig.~\ref{fig:scheme}.
Therefore, we split the set $\set{y}$ in $\set{\yp} = \set{(E,1), (N,1), (N,2)}$ and $\set{\yf} = \set{(E,2), (E,3), (N,3)}$, which is valid since the matrix whose entries are $\set{\ell^{\lambda}_{\yp}}$ is an identity matrix (see Eq.~\eqref{eq:allCLQD}).
The fields conjugated with the complete set of conservation laws, Eq.~\eqref{eq:effectiveIntensive}, are:
\begin{align}
	F_{E} & = \beta_{1} \, , & F_{\mathrm{u}} & = - \beta_{1} \mu_{1} \, , \; \text{and} & F_{\mathrm{d}} & = - \beta_{2} \mu_{2} \, ,
	\label{}
\end{align}
from which the reference potential of the state $n$, Eq.~\eqref{eq:PrefGG}, follows
\begin{equation}
	\P{gg}{n} = \Phi^{\mathrm{gg}} - \left[ - \beta_{1} E_{n} + \beta_{1} \mu_{1} N^{\mathrm{u}}_{n} + \beta_{2} \mu_{2} N^{\mathrm{d}}_{n} \right] \, .
	\label{eq:potentialQD}
\end{equation}
Instead, the fundamental forces, Eq.~\eqref{eq:fundForce}, are given by
\begin{equation}
	\begin{aligned}
		\mathcal{F}_{(E,2)} & = \beta_{1} - \beta_{2} \, , &
		\mathcal{F}_{(E,3)} & = \beta_{1} - \beta_{3} \, , \; \text{and} \\
		\mathcal{F}_{(N,3)} & = \beta_{3} \mu_{3} - \beta_{2} \mu_{2} \, ,
	\end{aligned}
	\label{eq:effAffQD}
\end{equation}
from which the reference affinities follow, Eq.~\eqref{eq:ArefGG}.
The first two forces drive the energy flowing into the first reservoir from the second and third ones, respectively, whereas the third force drives the electrons flowing from the third to the second reservoir.

\section{Conclusions}

In this paper, we presented a general method to construct DFTs for Markov jump processes.
The strategy to identify the fluctuating quantities which satisfy the DFT consists of splitting the EP in two by making use of a reference PMF.
The choice of the reference PMF is arbitrary for IFTs, but must solely depend on the driving protocol for DFTs.
Out of the infinite number of FTs that can be considered, we tried to select those that have interesting mathematical properties or that can be expressed in terms of physical quantities when the Markov jump process is complemented with a thermodynamic structure.
Tab.~\ref{tab:master} summarizes the terms of to the EP for each of our choices.
We also emphasized that the EP always satisfies an IFT but generically not a DFT.
Connections to information theory were also made by formulating a generalized Landauer principle.

We do not claim to have been exhaustive, and many other reference PMF{s} may be interesting.
We can mention at least two more interesting cases.
By considering the steady-state PMF which is obtained when removing some edges from the graph (but not all chords as in Sec.~\ref{sec:ste}), the marginal thermodynamic theory presented in Refs.~\cite{polettini17,polettini18} emerges.
One can also consider a reference PMF in between the microcanonical PMF, which takes no conserved quantity into account, and the generalized Gibbs one, which takes them all into account.
This happens for instance when only the obvious conserved quantities are accounted for, $\set{\X}$, as discussed in Ref.~\cite{bulnescuetara14}.
In this case, one uses the fields of a given reservoir to define the reference equilibrium potential
\begin{equation*}
	\Pref{n} = \Phi - \left[ S_{n} - {\textstyle\sum_{\lX}} f_{(\lX,1)} \deltaX^{\lX}_{n} \right] \, ,
\end{equation*}
where $\Phi$ is determined by the normalization.
The number of nonconservative forces appearing in $\vepr{nc}$ will be $\nof{\mathrm{y}} - \nof{\kappa}$.
However, in case additional conservation laws are present ($\nof{\lambda} > \nof{\kappa}$), some of these forces are dependent on others and their number will be larger than the minimal, $\nof{\mathrm{y}} - \nof{\lambda}$.

\acknowledgments

We thank A.~Wachtel and A.~Lazarescu for valuable feedback on the manuscript.\myFunding

\section*{Abbreviations}

The following abbreviations are used in this paper:\\
\noindent 
\begin{tabular}{@{}ll}
	DFT & detailed fluctuation theorem \\
	IFT & integral fluctuation theorem \\
	PMF & probability mass function \\
	EP & entropy production \\
	ME & master equation \\
	MGF & moment generating function \\
\end{tabular}

\appendix

\begin{widetext}
\section{Moment Generating Function Dynamics and Proofs of the FTs}
\label{sec:ftMasterProof}

We describe the moment generating function (MGF) technique that we use to prove the finite time DFTs \eqref{eq:dftMaster} \cite{rao18:shape}.

\subsection*{MGF Dynamics}

Let ${P_{t}(n,\deltaO)}$ be the joint probability of observing a trajectory ending in the state $n$ along which the change of a generic observable, $\obs$, is $\deltaO$.
The changes of $\obs$ along edges are denoted as $\set{\deltaO_{e}}$, whereas the changes due to time-dependent driving while in the state $n$ as $\dot{O}_{n}$.
In order to write an evolution equation for this probability, let us expand it as:
\begin{equation}
	P_{t+\de t}( n, \deltaO ) \simeq
	{\textstyle\sum_{e}} w_{e} \delta_{n,\t{e}} \, P_{t}\left( \o{e}, \deltaO - \deltaO_{e} - \dot{\obs}_{\o{e}} \de t \right) \de t + \left[ 1 - {\textstyle\sum_{e}} w_{e} \delta_{n,\o{e}} \de t \right] P_{t}( n, \deltaO - \dot{\obs}_{n} \de t) \, .
\end{equation}
The first term accounts for transitions leading to the state $n$ and completing the change of $\obs$, whereas the second describes the probability of completing the change of $\obs$ while dwelling in the state $n$ (and not leaving it).
When keeping only the linear term in $\de t$ and performing the limit $\de t \rightarrow 0$, we get:
\begin{equation}
	\dt P_{t}( n, \deltaO ) =
	{\textstyle\sum_{e}} w_{e} \delta_{n,\t{e}} \, P_{t}\left( \o{e}, \deltaO - \deltaO_{e} \right)
	- {\textstyle\sum_{e}} w_{e} \delta_{n,\o{e}} \, P_{t}( n, \deltaO ) 
	- \dot{\obs}_{n} \partial_{\deltaO} P_{t}( n, \deltaO ) \, .
	\label{eq:evoP(n,deltaO)}
\end{equation}
Rather than working with this differential equation, it is much more convenient to deal with the bilateral Laplace transform of $p_{t}( n, \deltaO )$, that is, the MGF up to a sign,
\begin{equation}
	\Lambda_{n,t}(\cf) := {\textstyle\int_{-\infty}^{\infty}} \de \, \deltaO \, \exp\left\{ - \cf \deltaO \right\} P_{t}( n, \deltaO ) \, ,
	\label{eq:mgf}
\end{equation}
since its evolution equation is akin to an ME, Eq.~\eqref{eq:MEgenerator}:
\begin{equation}
	\dt \Lambda_{n,t}(\cf) =
	{\textstyle\sum_{m}} W_{nm,t}(q) \, \Lambda_{m,t}(\cf) \, ,
	\label{eq:mgfDynam}
\end{equation}
where the \emph{biased rate matrix} reads
\begin{equation}
	W_{nm,t}(\cf) = {\textstyle\sum_{e}} w_{e} \left\{ \exp\left\{ - \cf \deltaO_{e} \right\} \delta_{n,\t{e}} \delta_{m,\o{e}} - \delta_{n,m} \delta_{m,\o{e}} \right\}
	- \cf \, \dot{\obs}_{n} \delta_{n,m} \, .
	\label{eq:tiltedGenerator}
\end{equation}
The field $q$ is usually referred to as {a} \emph{counting field}.
This equation is obtained by combining Eqs.~\eqref{eq:evoP(n,deltaO)} and \eqref{eq:mgf}, and its initial condition must be $\Lambda_{n,0}(\deltaO) = p_{n}(0)$.
Note that Eq.~\eqref{eq:mgfDynam} is not an ME, since ${\sum_{n}} \Lambda_{n,t}(\deltaO)$ is not conserved.

For later convenience, we recast Eq.~\eqref{eq:mgfDynam} into a bracket notation:
\begin{equation}
	\dt \ket{\Lambda_{t}( \cf )} = {\mathcal{W}}_{t}(\cf) \ket{\Lambda_{t}( \cf )} \, ,
	\label{eq:MEgfCompact}
\end{equation}
and we proceed to prove a preliminary result.
A formal solution of Eq.~\eqref{eq:mgfDynam} is $\ket{\Lambda_{t}(\cf)} = {\mathcal{U}}_{t}(\cf) \, \ket{P(0)}$, where the time-evolution operator reads $\mathcal{U}_{t}(\cf) = \mathsf{T}_{+} \exp \int_{0}^{t} \de \tau \, {\mathcal{W}}_{\tau}(\cf)$, $\mathsf{T}_{+}$ being the time-ordering operator.
We clearly have $\dt \mathcal{U}_{t}(\cf) = {\mathcal{W}}_{t}(\cf) \, \mathcal{U}_{t}(\cf)$.
Let us now consider the following transformed evolution operator:
\begin{equation}
	\tilde{\mathcal{U}}_{t}(\cf) := \mathcal{X}^{-1}_{t} \mathcal{U}_{t}(\cf) {\mathcal{X}}_{0} \, ,
	\label{eq:transformedUdef}
\end{equation}
${\mathcal{X}}_{t}$ being a generic time-dependent invertible operator.
Its dynamics is ruled by the following biased stochastic dynamics:
\begin{equation}
	\dt \tilde{\mathcal{U}}_{t}(\cf)
	= \dt \mathcal{X}^{-1}_{t} \mathcal{U}_{t}(\cf) {\mathcal{X}}_{0} + \mathcal{X}^{-1}_{t} \dt \mathcal{U}_{t}(\cf) {\mathcal{X}}_{0}
	= \left\{ \dt \mathcal{X}^{-1}_{t} {\mathcal{X}}_{t} + \mathcal{X}^{-1}_{t} {\mathcal{W}}_{t}(\cf) {\mathcal{X}}_{t} \right\} \tilde{\mathcal{U}}_{t}(\cf)
	\equiv \tilde{\mathcal{W}}_{t}(\cf) \, \tilde{\mathcal{U}}_{t}(\cf) \, ,
	\label{eq:transformedUdynamics}
\end{equation}
which allows us to conclude that the transformed time-evolution operator is given by
\begin{equation}
	\tilde{\mathcal{U}}(\cf) = \mathsf{T}_{+} \exp {\textstyle\int_{0}^{t}} \de \tau \, \tilde{\mathcal{W}}_{\tau}(\cf) \, .
	\label{eq:transformedU}
\end{equation}

From Eqs.~\eqref{eq:transformedUdef}, \eqref{eq:transformedUdynamics}, and \eqref{eq:transformedU}, we deduce that
\begin{equation}
	\mathcal{X}^{-1}_{t} \mathcal{U}_{t}(\cf) {\mathcal{X}}_{0}
	= \mathsf{T}_{+} \exp {\textstyle\int_{0}^{t}} \de \tau \, \left[ \de_{\tau} \mathcal{X}^{-1}_{\tau} {\mathcal{X}}_{\tau} + \mathcal{X}^{-1}_{\tau} {\mathcal{W}}_{\tau}(\cf) {\mathcal{X}}_{\tau} \right] \, .
	\label{eq:magic}
\end{equation}

\subsection*{Proof of the DFT}

To prove the DFT \eqref{eq:dftMaster}, we briefly recall its two assumptions:
\emph{(i)} the reference PMF depends on time solely via the protocol function;
\emph{(ii)} for both the forward and backward processes, the system is initially prepared in a reference PMF.
Let ${P_{t}(n,\vep{d},\vep{nc})}$ be the joint probability of observing a trajectory ending in the state $n$ along which the driving contribution is $\vep{d}$, while the nonconservative one is $\vep{nc}$.
The above probabilities, one for each $n$, are stacked in the ket $\ket{P_{t}(\vep{d}, \vep{nc})}$.
The time evolution of the related MGF,
\begin{equation}
	\ket{\Lambda_{t} (\cf_{\mathrm{d}},\cf_{\mathrm{nc}})} :=
	{\textstyle\int^{\infty}_{-\infty}} \de \vep{d} \de \vep{nc}
	\exp\left\{ -\cf_{\mathrm{d}} \vep{d} - \cf_{\mathrm{nc}} \vep{nc} \right\}
	\ket{P_{t}(\vep{d}, \vep{nc})} \, ,
	\label{eq:mgfDNC}
\end{equation}
is ruled by the biased stochastic dynamics, Eq.~\eqref{eq:mgfDynam},
\begin{equation}
	\dt \ket{\Lambda_{t}(\cf_{\mathrm{d}},\cf_{\mathrm{nc}})} =
	{\mathcal{W}}_{t}(\cf_{\mathrm{d}},\cf_{\mathrm{nc}}) \ket{\Lambda_{t}(\cf_{\mathrm{d}},\cf_{\mathrm{nc}})} \, ,
	\label{eq:biasedME}
\end{equation}
where the entries of the biased generator are given by
\begin{equation}
	W_{nm}(\cf_{\mathrm{d}},\cf_{\mathrm{nc}})
	= {\textstyle\sum_{e}} w_{e} \big\{ \exp\left\{ - \cf_{\mathrm{nc}} \Aref{} \right\} \delta_{n,\t{e}} \delta_{m,\o{e}}
	- \delta_{n,m} \delta_{m,\o{e}} \big\} - \cf_{\mathrm{d}} \dt \psi_{m} \delta_{n,m} \, .
	\label{eq:biasedGeneratorBis}
\end{equation}
Using the definition of reference affinity, Eq.~\eqref{eq:Aref}, one can see that the rate matrix satisfies the following symmetry:
\begin{equation}
	{\mathcal{W}}_{t}\transpose(\cf_{\mathrm{d}},\cf_{\mathrm{nc}}) = \mathcal{P}_{t}^{-1} \, \mathcal{W}_{t}(\cf_{\mathrm{d}}, 1 - \cf_{\mathrm{nc}}) \, \mathcal{P}_{t} \, ,
	\label{eq:symmetryOper}
\end{equation}
where the entries of $\mathcal{P}_{t}$ are given by
\begin{equation}
	\mathcal{P}_{nm,t} := \exp\left\{ - \Pref{m}(\pi_{t}) \right\} \delta_{n,m} \, ,
	\label{eq:B}
\end{equation}
and ``$\,\transpose\,$'' denotes the transposition.
Additionally, the initial condition is given by the reference PMF:
\begin{equation}
	\ket{\Lambda_{0}(\cf_{\mathrm{d}},\cf_{\mathrm{nc}})}
	= \ket{\pref{0}}
	= \mathcal{P}_{0} \ket{1} \, .
	\label{}
\end{equation}
$\ket{1}$ denotes the vector in the state space whose entries are all equal to one.

Using the formal solution of Eq.~\eqref{eq:biasedME}, the MGF of ${P_{t}(\vep{d}, \vep{nc})}$ can be written as:
\begin{equation}
	\Lambda_{t}(\cf_{\mathrm{d}},\cf_{\mathrm{nc}})
	=  \braket{1|\Lambda_{t}(\cf_{\mathrm{d}},\cf_{\mathrm{nc}})} 
	= \braket{1| \mathcal{U}_{t}(\cf_{\mathrm{d}},\cf_{\mathrm{nc}}) \mathcal{P}_{0} | 1}
	= \braket{1| \mathcal{P}_{t} \mathcal{P}_{t}^{-1} \, \mathcal{U}_{t}(\cf_{\mathrm{d}},\cf_{\mathrm{nc}}) \, \mathcal{P}_{0} | 1} ,\
	\label{eq:proofFirst}
\end{equation}
where $\mathcal{U}_{t}(\cf_{\mathrm{d}},\cf_{\mathrm{nc}})$ is the related time-evolution operator.
Using the relation in Eq.~\eqref{eq:magic}, the last term can be recast into
\begin{equation}
	{\Lambda_{t}(\cf_{\mathrm{d}},\cf_{\mathrm{nc}})}
	= \braket{\pref{t} | \mathsf{T}_{+} \exp\left\{ {\textstyle\int_{0}^{t}} \de \tau \, \left[ \de_{\tau} \mathcal{P}_{\tau}^{-1} \mathcal{P}_{\tau} + \mathcal{P}_{\tau}^{-1} \, {\mathcal{W}}_{\tau}(\cf_{\mathrm{d}},\cf_{\mathrm{nc}}) \, \mathcal{P}_{\tau} \right] \right\} |1} \, .
	\label{}
\end{equation}
Since $\de_{\tau} \mathcal{P}_{\tau}^{-1} \mathcal{P}_{\tau} = \diag\left\{ \de_{\tau} \Pref{n} \right\}$, the first term in square brackets can be added to the diagonal entries of the second term, thus giving
\begin{equation}
	{\Lambda_{t}(\cf_{\mathrm{d}},\cf_{\mathrm{nc}})}
	= \braket{\pref{t} | \mathsf{T}_{+} \exp\left\{ {\textstyle\int_{0}^{t}} \de \tau \, \left[ \mathcal{P}_{\tau}^{-1} \, {\mathcal{W}}_{\tau}(\cf_{\mathrm{d}} - 1,\cf_{\mathrm{nc}}) \, \mathcal{P}_{\tau} \right] \right\} |1} \, .
	\label{}
\end{equation}
The symmetry \eqref{eq:symmetryOper} allows us to recast the latter into
\begin{equation}
	{\Lambda_{t}(\cf_{\mathrm{d}},\cf_{\mathrm{nc}})}
	= \braket{\pref{t} | \mathsf{T}_{+} \exp\left\{ {\textstyle\int_{0}^{t}} \de \tau \, \mathcal{W}\transpose_{\tau}\left( \cf_{\mathrm{d}} - 1, 1 - \cf_{\mathrm{nc}} \right) \right\} |1} \, .
	\label{eq:intermediario}
\end{equation}
The crucial step comes as we time-reverse the integration variable: $\tau \rightarrow t - \tau$.
Accordingly, the time-ordering operator, $\mathsf{T}_{+} $, becomes an anti-time-ordering one, $\mathsf{T}_{-}$, while the diagonal entries of the biased generator become
\begin{equation}
	W_{mm,t-\tau}(\cf_{\mathrm{d}},\cf_{\mathrm{nc}})
	= - {\textstyle\sum_{e}} w_{e}(\pi_{t-\tau}) \, \delta_{m,\o{e}} - \cf_{\mathrm{d}} \, \de_{t-\tau} \Pref{m}(\pi_{t-\tau})
	= - {\textstyle\sum_{e}} w_{e}(\pi^{\dagger}_{\tau}) \, \delta_{m,\o{e}} + \cf_{\mathrm{d}} \, \de_{\tau} \Pref{m}(\pi^{\dagger}_{\tau}) \, ,
	\label{}
\end{equation}
from which we conclude that
\begin{equation}
	W_{nm,t-\tau}(\cf_{\mathrm{d}},\cf_{\mathrm{nc}})
	= W^{\dagger}_{nm,\tau}(-\cf_{\mathrm{d}},\cf_{\mathrm{nc}}) \, .
	\label{}
\end{equation}
Crucially, the assumption that $\Pref{n}$ depends on time via $\pi_{\tau}$ ensures that ${\mathcal{W}}^{\dagger}_{\tau}(\cf_{\mathrm{d}},\cf_{\mathrm{nc}})$ can be regarded as the biased generator of the dynamics subject to the time-reversed protocol (\emph{i.e.}, the dynamics of the backward process).
If we considered an arbitrary $\pref{n}$ (\emph{i.e.}, the forward process would start from an arbitrary PMF), then ${\mathcal{W}}^{\dagger}_{\tau}(\cf_{\mathrm{d}},\cf_{\mathrm{nc}})$ would be the rate matrix of the time-reversed stochastic dynamics:
\begin{equation}
	0 = {\textstyle\sum_{m}} \left[ \delta_{nm} \de_{t-\tau} - W_{nm}(\pi_{t-\tau}) \right] p_{m}
	= {\textstyle\sum_{m}} \left[ - \delta_{nm} \de_{\tau} - W_{nm}(\pi^{\dagger}_{\tau}) \right] p_{m} \, ,
	\label{eq:trME}
\end{equation}
which is unphysical.
Equation~\eqref{eq:intermediario} thus becomes
\begin{equation}
	{\Lambda_{t}(\cf_{\mathrm{d}},\cf_{\mathrm{nc}})}
	= \braket{\pref{t} | \mathsf{T}_{-} \exp\left\{ {\textstyle\int_{0}^{t}} \de \tau \, {\mathcal{W}_{\tau}^{\dagger}}\transpose \left( 1 - \cf_{\mathrm{d}},1 - \cf_{\mathrm{nc}} \right) \right\} |1} \, .
	\label{}
\end{equation}
Upon a global transposition, we can write
\begin{equation}
	{\Lambda_{t}(\cf_{\mathrm{d}},\cf_{\mathrm{nc}})} = \braket{1 | \mathsf{T}_{+} \exp\left\{ {\textstyle\int_{0}^{t}} \de \tau \, {\mathcal{W}_{\tau}^{\dagger}} \left( 1 - \cf_{\mathrm{d}},1 - \cf_{\mathrm{nc}} \right) \right\} |\pref{t}} \, ,
\end{equation}
where we also used the relationship between transposition and time-ordering
\begin{equation}
	\mathsf{T}_{+} \left( {\textstyle\prod_{i}} A\transpose_{t_{i}} \right) = \left( \mathsf{T}_{-} {\textstyle\prod_{i}} A_{t_{i}} \right)\transpose \, ,
	\label{}
\end{equation}
in which $A_{t}$ is a generic operator.
From the last expression, we readily obtain the symmetry that we are looking for:
\begin{equation}
	{\Lambda_{t}(\cf_{\mathrm{d}},\cf_{\mathrm{nc}})}
	= \Lambda^{\dagger}_{t}\left( 1 - \cf_{\mathrm{d}},1 - \cf_{\mathrm{nc}} \right) \, ,
	\label{eq:proofLast}
\end{equation}
where $\Lambda^{\dagger}_{t}\left( \cf_{\mathrm{d}}, \cf_{\mathrm{nc}} \right)$ is the MGF of $P^{\dagger}_{t}(\vep{d}, \vep{nc})$.
Indeed, its inverse Laplace transform gives the DFT in Eq.~\eqref{eq:dftMaster}.

\subsection*{Proof of the DFT for the Sum of Driving and Nonconservative EP}

Let us define $\vep{s} := \vep{d}+\vep{nc}$ as the sum of the driving and nonconservative EP contributions.
A straightforward calculation leads from \eqref{eq:dftMaster} to the DFT for $\vep{s}$, Eq.~\eqref{eq:dftMasterSum}:
\begin{equation}
	\begin{split}
		P_{t}(\vep{s})
		& = {\textstyle\int} \de \vep{d} \de \vep{nc} \, P_{t}(\vep{d},\vep{nc}) \, \delta\left( \vep{s} - \vep{d} - \vep{nc} \right)
		= {\textstyle\int} \de \vep{d} \, P_{t}(\vep{d},\vep{s}-\vep{d}) \\
		& = \exp{\vep{s}} \, {\textstyle\int} \de \vep{d} \, P^{\dagger}_{t}(-\vep{d},\vep{d}-\vep{s})
		= P^{\dagger}_{t}(-\vep{s}) \, \exp{\vep{s}} \, .
	\end{split}
	\label{}
\end{equation}

\subsection*{Proof of the IFT}

We now prove the IFT \eqref{eq:iftMaster} using the MGF technique developed in Ref.~\cite{esposito07}.
We have already mentioned that the dynamics \eqref{eq:biasedME} does not describe a stochastic process, since the normalization is not preserved.
However, for $\cf_{\mathrm{d}} = \cf_{\mathrm{nc}} = 1$, the biased generator \eqref{eq:biasedGeneratorBis} can be written as:
\begin{equation}
	W_{nm}(1,1)
	= \Big[ {\textstyle\sum_{e}} w_{e} \pref{\o{e}} \big\{ \delta_{n,\o{e}} \delta_{m,\t{e}}
	- \delta_{n,m} \delta_{m,\o{e}} \big\} + \dt \pref{n} \delta_{n,m} \Big] \frac{1}{\pref{m}} \, ,
	\label{eq:gen11}
\end{equation}
from which it readily follows that
\begin{equation}
	\dt \ket{\pref{}} = \mathcal{W}(1,1) \ket{\pref{}} \, ,
	\label{}
\end{equation}
\emph{viz.} $\pref{n}$ is the solution of the biased dynamics \eqref{eq:biasedME} for $\cf_{\mathrm{d}} = \cf_{\mathrm{nc}} = 1$.
The normalization condition thus demands that
\begin{equation}
	1 = \braket{1|\Lambda_{t} (1,1)}
	= {\textstyle\int^{\infty}_{-\infty}} \de \vep{d} \de \vep{nc} \exp\left\{ - \vep{d} - \vep{nc} \right\} \braket{1|P_{t}(\vep{d}, \vep{nc})}
	\equiv \avef{\exp\left\{ - \vep{d} - \vep{nc} \right\}} \, ,
	\label{}
\end{equation}
which is the IFT in Eq.~\eqref{eq:iftMaster}.
Note that we do not assume any specific property for $\pref{n}$ in this context.

\section{Alternative Proofs of the DFT}
\label{sec:ftMasterProof2}

We here show two alternative proofs of the DFT \eqref{eq:dftMaster} which rely on the involution property \eqref{eq:involution}.
For the nonadiabatic contribution, this property can be proved as follows.
By time-reversing Eq.~\eqref{eq:epMasterNC}, $\tau \rightarrow  t - \tau$, we obtain
\begin{equation}
	\vep{nc}[\trj;\prt] = {\textstyle\int_{0}^{t}} \de \tau \, \Aref{}(\pi_{\tau}) \, j^{e}(\tau)
	= {\textstyle\int_{0}^{t}} \de \tau \, \Aref{}(\pi_{t-\tau}) \, j^{e}(t-\tau) \, .
	\label{}
\end{equation}
Since $\Aref{}$ is solely determined by the state of protocol at each instant of time, the reference affinities correspond to those of the backward process, $\Aref{}(\pi_{t-\tau}) = \Aref{}(\pi^{\dagger}_{\tau})$.
Using the property that $j^{e}(t-\tau) = j^{\dagger\, - e}(\tau)$, see Eq.~\eqref{eq:trCrr}, and $\Aref{} = - \Aref{-}$, we finally obtain
\begin{equation}
	\vep{nc}[\trj;\prt]
	= - {\textstyle\int_{0}^{t}} \de \tau \, \Aref{}(\pi^{\dagger}_{\tau}) \, {j^{\dagger\, e}}(\tau)
	= - \vep{nc}[\trj^{\dagger};\prt^{\dagger}] \, .
	\label{}
\end{equation}
Concerning the driving contribution, Eq.~\eqref{eq:epMasterD}, we obtain
\begin{equation}
	\vep{d}[\trj;\prt] = {\int_{0}^{t}} \de \tau \at{\left[ \de_{\tau} \Pref{n}(\pi_{\tau}) \right]}{n=n_{\tau}}
	= {\int_{0}^{t}} \de \tau \at{\left[ - \de_{\tau} \Pref{n}(\pi_{t-\tau}) \right]}{n=n_{t-\tau}} \, .
	\label{}
\end{equation}
It is here again crucial that $\Pref{n}$ depends solely on the protocol value, so that $\Pref{n}(\pi_{t-\tau}) = \Pref{n}(\pi^{\dagger}_{\tau})$.
Therefore,
\begin{equation}
	\vep{d}[\trj;\prt]
	= - {\int_{0}^{t}} \de \tau \at{\left[ \de_{\tau} \Pref{n}(\pi^{\dagger}_{\tau}) \right]}{n=n^{\dagger}_{\tau}}
	= - \vep{d}[\trj^{\dagger};\prt^{\dagger}] \, .
	\label{}
\end{equation}

\subsection*{Alternative Proof 1}
\label{sec:ftAlt1}

Inspired by Ref.~\cite{garcia-garcia10}, we here use an alternative approach to derive the symmetry of the MGF which underlies our DFT, Eq.~\eqref{eq:proofLast}.
In terms of trajectory probabilities, the MGF \eqref{eq:mgfDNC} can be written as:
\begin{equation}
	\Lambda_{t} (\cf_{\mathrm{d}},\cf_{\mathrm{nc}})
	= {\textstyle\int \mathfrak{D}\trj} \, \mathfrak{P}[\trj;\prt] \, \pref{n_{0}}(\pi_{0}) \,
	\exp\left\{ - \cf_{\mathrm{d}} \vep{d}[\trj;\prt] - \cf_{\mathrm{nc}} \vep{nc}[\trj;\prt] \right\} \, .
	\label{}
\end{equation}
Using the relation between the EP contributions and the stochastic trajectories in forward and backward processes, Eq.~\eqref{eq:vepFluctuating}, we can recast the MGF into
\begin{equation}
	{\Lambda_{t} (\cf_{\mathrm{d}},\cf_{\mathrm{nc}})}
	= {\textstyle\int \mathfrak{D}\trj} \, \mathfrak{P}[\trj^{\dagger};\prt^{\dagger}] \, \pref{n_{t}}(\pi_{t}) \,
	\exp\left\{ \left( 1 - \cf_{\mathrm{d}} \right) \vep{d}[\trj;\prt] + \left( 1 - \cf_{\mathrm{nc}} \right) \vep{nc}[\trj;\prt] \right\} \, ,
	\label{}
\end{equation}
so that using the property of involution, Eq.~\eqref{eq:involution}, we get
\begin{equation}
	{\Lambda_{t} (\cf_{\mathrm{d}},\cf_{\mathrm{nc}})}
	= {\textstyle\int \mathfrak{D}\trj} \, \mathfrak{P}[\trj^{\dagger};\prt^{\dagger}] \, \pref{n_{t}}(\pi_{t}) \,
	\exp\left\{ - \left( 1 - \cf_{\mathrm{d}} \right) \vep{d}[\trj^{\dagger};\prt^{\dagger}] - \left( 1 - \cf_{\mathrm{nc}} \right) \vep{nc}[\trj^{\dagger};\prt^{\dagger}] \right\} \, .
	\label{}
\end{equation}
Hence, changing and renaming the integration variable, $\trj \rightarrow \trj^{\dagger}$, and using the fact that the Jacobian determinant of this transformation is one, we finally get
\begin{equation}
	\hspace{-1em}
	{\Lambda_{t} (\cf_{\mathrm{d}},\cf_{\mathrm{nc}})}
	= {\textstyle\int \mathfrak{D}\trj} \, \mathfrak{P}[\trj;\prt^{\dagger}] \, \pref{n_{t}}(\pi_{t}) \,
	\exp\left\{ - \left( 1 - \cf_{\mathrm{d}} \right) \vep{d}[\trj;\prt^{\dagger}] - \left( 1 - \cf_{\mathrm{nc}} \right) \vep{nc}[\trj;\prt^{\dagger}] \right\}
	= \Lambda^{\dagger}_{t}\left( 1 - \cf_{\mathrm{d}},1 - \cf_{\mathrm{nc}} \right) \, ,
	\label{}
\end{equation}
which proves Eq.~\eqref{eq:proofLast}.
With respect to the previous proof, this one is based on Eq.~\eqref{eq:vepFluctuating} and on the property of involution, which follow from the specifications of forward and backward processes.

\subsection*{Alternative Proof 2}
\label{sec:ftAlt2}

The joint probability distribution $P_{t}(\vep{d},\vep{nc})$ written in terms of trajectory probabilities, Eq.~\eqref{eq:probTrajectory}, reads
\begin{equation}
	P_{t}(\vep{d},\vep{nc})
	= {\textstyle\int \mathfrak{D}\trj} \, \mathfrak{P}[\trj;\prt] \, \pref{n_{0}}(\pi_{0}) \, \delta\left( \vep{d}[\trj;\prt] - \vep{d} \right) \, \delta\left( \vep{nc}[\trj;\prt] - \vep{nc} \right) \, .
	\label{}
\end{equation}
Using Eq.~\eqref{eq:vepFluctuating} and then the involution property \eqref{eq:involution}, we finally obtain the DFT \eqref{eq:dftMaster}:
\begin{equation}
	\begin{split}
		P_{t}(\vep{d},\vep{nc})
		& = \exp\left\{ \vep{d} + \vep{nc} \right\} {\textstyle\int \mathfrak{D}\trj} \, \mathfrak{P}[\trj^{\dagger};\prt^{\dagger}] \, \pref{n_{t}}(\pi_{t}) \, \delta\left( \vep{d}[\trj;\prt] - \vep{d} \right) \, \delta\left( \vep{nc}[\trj;\prt] - \vep{nc} \right) \\
		& = \exp\left\{ \vep{d} + \vep{nc} \right\} {\textstyle\int \mathfrak{D}\trj} \, \mathfrak{P}[\trj^{\dagger};\prt^{\dagger}] \, \pref{n_{t}}(\pi_{t}) \, \delta\left( - \vep{d}[\trj^{\dagger};\prt^{\dagger}] - \vep{d} \right) \, \delta\left( - \vep{nc}[\trj^{\dagger};\prt^{\dagger}] - \vep{nc} \right) \\
		& = \exp\left\{ \vep{d} + \vep{nc} \right\} P_{t}^{\dagger}(-\vep{d},-\vep{nc}) \, .
	\end{split}
\end{equation}

\section{Adiabatic and Nonadiabatic Contributions}
\label{sec:ftMasterProofAdNonad}

We now prove that both the adiabatic and nonadiabatic EP rates are non-negative.
Concerning the adiabatic contribution, using the log-inequality, $- \ln x \ge 1 - x$, one obtains
\begin{equation}
	\hspace{-2.4em}
	\vepr{a}
	=  {\sum_{e}} w_{e} p_{\o{e}} \ln \frac{w_{e} \pss{\o{e}}}{w_{-e} \pss{\o{-e}}}
	\ge {\sum_{e}} w_{e} p_{\o{e}} \Bigg[ 1 - \frac{w_{-e} \pss{\o{-e}}}{w_{e} \pss{\o{e}}} \Bigg]
	= {\sum_{e}} \left[ w_{e} \pss{\o{e}} - w_{-e} \pss{\o{-e}} \right] \frac{p_{\o{e}}}{\pss{\o{e}}}
	= {\sum_{e,n}} D^{e}_{n} w_{e} \pss{\o{e}} \left[ - \frac{p_{n}}{\pss{n}} \right]
	= 0 \, .
	\label{}
\end{equation}
The last equality follows from the definition of steady-state PMF, Eq.~\eqref{eq:prefSS}.
For the nonadiabatic, instead, using the same inequality and similar algebraic steps, one obtains:
\begin{equation}
	\begin{split}
		\vepr{na}
		& = {\sum_{e}} w_{e} p_{\o{e}} \ln \frac{p_{\o{e}} \pss{\o{-e}}}{\pss{\o{e}} p_{\o{-e}}}
		\ge {\sum_{e}} w_{e} p_{\o{e}} \Bigg[ 1 - \frac{\pss{\o{e}} p_{\o{-e}}}{p_{\o{e}} \pss{\o{-e}}} \Bigg]
		= {\sum_{e}} \left[ w_{e} \pss{\o{e}} - w_{-e} \pss{\o{-e}} \right] \frac{p_{\o{e}}}{\pss{\o{e}}}
		= 0 \, .
	\end{split}
	\label{}
\end{equation}

\section{Proofs of the DFTs for the Adiabatic and Driving EP Contributions}
\label{sec:dftAdNonadProof}

We here prove the DFTs in Eqs.~\eqref{eq:DFTa} and \eqref{eq:DFTd} using the same MGF technique described in App.~\ref{sec:ftMasterProof}.

\subsection*{Proof of the DFT for the Adiabatic Contribution}

The biased generator ruling the sole adiabatic term reads:
\begin{equation}
	W_{nm}(\cf_{\mathrm{a}})
	= {\textstyle\sum_{e}} w_{e}
	\left\{ \exp\left\{ - \cf_{\mathrm{a}} \A{ss}{} \right\} \delta_{n,\t{e}} \delta_{m,\o{e}} - \delta_{n,m} \delta_{m,\o{e}} \right\} \, .
\end{equation}
It satisfies the following symmetry:
\begin{equation}
	\mathcal{W}(\cf_{\mathrm{a}}) = \hat{\mathcal{W}}(1-\cf_{\mathrm{a}}) \, ,
	\label{eq:symmetryWa}
\end{equation}
where $\hat{\mathcal{W}}(\cf_{\mathrm{a}})$ is the biased generator of the fictitious dynamics ruled by the rates in Eq.~\eqref{eq:fictRates}.
Crucially, $\pss{n}$ is also the steady state of this dynamics:
\begin{equation}
	{\textstyle\sum_{e}} D^{n}_{e} \, \hat{w}_{e} \pss{\o{e}}
	= {\textstyle\sum_{m}\sum_{e}} \hat{w}_{e} \left\{ \delta_{n,\t{e}} \delta_{m,\o{e}} - \delta_{n,m} \delta_{m,\o{e}} \right\} \pss{m}
	= 0 \, , \quad \forall \, n \, .
	\label{}
\end{equation}
This fact guarantees that the escape rates of the fictitious dynamics coincide with those of the original ones:
\begin{equation}
	- {\textstyle\sum_{e}} \hat{w}_{e} \delta_{n,m} \delta_{m,\o{e}} = 
	- {\textstyle\sum_{e}} w_{e} \delta_{n,m} \delta_{m,\o{e}}
	\, , \quad \forall \, n \, .
	\label{}
\end{equation}
We can now proceed to prove the FT \eqref{eq:DFTa}:
\begin{equation}
	\Lambda_{t}(\cf_{\mathrm{a}})
	= \braket{1|\Lambda_{t}(\cf_{\mathrm{a}})} 
	= \braket{1| \mathcal{U}_{t}(\cf_{\mathrm{a}}) | p}
	= \braket{1| \mathsf{T}_{+} \exp\left\{ {\textstyle\int_{0}^{t}} \de \tau \, {\mathcal{W}}_{\tau}(\cf_{\mathrm{a}}) \right\} | p}
	= \braket{1| \mathsf{T}_{+} \exp\left\{ {\textstyle\int_{0}^{t}} \de \tau \, \hat{\mathcal{W}}_{\tau}(1-\cf_{\mathrm{a}}) \right\} | p} \, .
	\label{eq:proofFirstA}
\end{equation}
In the last equality, we made use of the symmetry in Eq.~\eqref{eq:symmetryWa}.
Following the same mathematical steps backward, we readily get
\begin{equation}
	\Lambda_{t}(\cf_{\mathrm{a}})
	= \hat{\Lambda}_{t}(1-\cf_{\mathrm{a}}) \, ,
	\label{eq:proofSecondA}
\end{equation}
from which the DFT in Eq.~\eqref{eq:DFTa} ensues.

\subsection*{Proof of the DFT for the Driving Contribution}

Concerning the DFT of the driving term, Eq.~\eqref{eq:DFTd}, the generator of the related biased dynamics reads:
\begin{equation}
	W_{nm}(\cf_{\mathrm{d}})
	= {\textstyle\sum_{e}} w_{e} \left\{ \delta_{n,\t{e}} \delta_{m,\o{e}} - \delta_{n,m} \delta_{m,\o{e}} \right\} - \cf_{\mathrm{d}} \de_{t} \P{ss}{m} \delta_{n,m} \, ,
\end{equation}
and it satisfies the following symmetry:
\begin{equation}
	\hat{\mathcal{W}}_{t}\transpose(\cf_{\mathrm{d}},\cf_{\mathrm{nc}}) = \mathcal{P}_{t}^{-1} \, \mathcal{W}_{t}(\cf_{\mathrm{d}}, 1 - \cf_{\mathrm{nc}}) \, \mathcal{P}_{t} \, ,
	\label{eq:symmetryWd}
\end{equation}
where $\mathcal{P}_{t} := \diag \left\{ \exp - \P{ss}{m} \right\}$.
The finite-time DFT ensues when following the mathematical steps of the main proof and using Eq.~\eqref{eq:symmetryWd} at the step at Eq.~\eqref{eq:intermediario}.
\end{widetext}

\bibliography{indiceLocale}

\end{document}